\documentclass[fleqn,usenatbib]{mnras}
\usepackage{amsmath}
\usepackage{xspace}
\usepackage{mathptmx}
\usepackage{txfonts}
\usepackage{xcolor}

\usepackage[T1]{fontenc}
\usepackage{ae,aecompl}

\usepackage{graphicx}	% Including figure files
\usepackage{amssymb}	% Extra maths symbols
\usepackage{booktabs}
\usepackage{wasysym}
\usepackage{ulem}
\usepackage{pdflscape}
\usepackage{soul}

\let\oldhref\href
\renewcommand{\href}[2]{\oldhref{#1}{\hbox{#2}}}

\definecolor{colorl1}{RGB}{0, 51, 153}
\definecolor{colorl2}{RGB}{153, 0, 0}
\definecolor{colorl3}{RGB}{179, 179, 0}
\definecolor{colorl4}{RGB}{51, 102, 0}

\definecolor{colorw1}{RGB}{51, 102, 255}
\definecolor{colorw2}{RGB}{255, 51, 0}
\definecolor{colorw3}{RGB}{255, 214, 51}
\definecolor{colorw4}{RGB}{51, 204, 51}
\definecolor{seagreen}{rgb}{0.190, 0.525, 0.361} %added by Mario; remove if you need to

\newcommand{\ahf}{\texttt{AHF}}

\newcommand{\hMpc}{{\ifmmode{h^{-1}{\mathrm Mpc}}\else{$h^{-1}$Mpc}\fi}}
\newcommand{\Mpc}{{\ifmmode{{\mathrm Mpc}}\else{Mpc}\fi}}
\newcommand{\hkpc}{{\ifmmode{h^{-1}{\mathrm kpc}}\else{$h^{-1}$kpc}\fi}}
\newcommand{\kpc}{{\ifmmode{ {\mathrm kpc}}\else{{\mathrm kpc}}\fi}}
\newcommand{\kms}{{\ifmmode{ {\mathrm km\,s^{-1}}}\else{ ${\mathrm km\,s^{-1}}$}\fi}}
\newcommand{\hMsun}{{\ifmmode{h^{-1}{\mathrm {M_{\astrosun}}}}\else{$h^{-1}{\mathrm{M_{\astrosun}}}$}\fi}}
\newcommand{\Msun}{{\ifmmode{{\mathrm M}_{\astrosun}}\else{${\mathrm M}_{\astrosun}$}\fi}}
\newcommand{\Mhalo}{{\ifmmode{M_{\mathrm{halo}}}\else{$M_{\mathrm{halo}}$}\fi}}
\newcommand{\Rvir}{{\ifmmode{R_{\mathrm vir}}\else{$R_{\mathrm vir}$}\fi}}
\newcommand{\Rtwohun}{{\ifmmode{R_{200}}\else{$R_{200}$}\fi}}
\newcommand{\Mvir}{{\ifmmode{M_{\mathrm vir}}\else{$M_{\mathrm vir}$}\fi}}
\newcommand{\Mtwohun}{{\ifmmode{M_{200}}\else{$M_{200}$}\fi}}
\newcommand{\Nvir}{{\ifmmode{N_{\mathrm vir}}\else{$N_{\mathrm vir}$}\fi}}
\newcommand{\Mstar}{{\ifmmode{M_\star}\else{$M_\star$}\fi}}
\newcommand{\Vrot}{{\ifmmode{V_{\mathrm rot}}\else{$V_{\mathrm rot}$}\fi}}
\newcommand{\Reff}{{\ifmmode{R_{\mathrm e}}\else{$R_{\mathrm e}$}\fi}}
\newcommand{\Sigmae}{{\ifmmode{\Sigma_{\mathrm e}}\else{$\Sigma_{\mathrm e}$}\fi}}
\newcommand{\logmstar}{{\ifmmode \log(M_\star/\mathrm{M}_\odot) \else $\log(M_\star/\mathrm{M}_\odot)$ \fi}}
\newcommand{\ltsima}{$\; \buildrel < \over \sim \;$}
\newcommand{\gtsima}{$\; \buildrel > \over \sim \;$}
\newcommand{\lsim}{\lower.5ex\hbox{\ltsima}}
\newcommand{\gsim}{\lower.5ex\hbox{\gtsima}}

\def\lesssim{\mathrel{\hbox{\rlap{\hbox{\lower4pt\hbox{$\sim$}}}\hbox{$<$}}}}
\def\gtrsim{\mathrel{\hbox{\rlap{\hbox{\lower4pt\hbox{$\sim$}}}\hbox{$>$}}}}

\newcommand{\beq}{\begin{equation}}
\newcommand{\eeq}{\end{equation}}
\def\beqa{\begin{eqnarray}}
\def\eeqa{\end{eqnarray}}
\def\LCDM{\ensuremath{\Lambda}CDM}

\def\head{ \vbox to 0pt{\vss \hbox to 0pt{\hskip 440pt\mathrm
      LA-UR-10-07069\hss} \vskip 25pt}}

\def \kms {\ifmmode\,\mathrm{km\,s^{-1}}\else$\,\mathrm{km\,s^{-1}}$\fi}
\def \kpc {\ifmmode{\,\mathrm{kpc}}\else${\mathrm {kpc}}$\fi}  
\def \hkpc {\ifmmode{h^{-1}\mathrm{kpc}}\else${h^{-1}\mathrm{kpc}}$\fi}  
\def \hMpc {\ifmmode{h^{-1}\mathrm{Mpc}}\else${h^{-1}\mathrm{Mpc}}$\fi}  
\def \Mpc {\ifmmode{\mathrm{Mpc}}\else${\mathrm{Mpc}}$\fi}  
\def \Msun {\ifmmode{\mathrm{M}}_{\astrosun}\else${\mathrm{M}}_{\astrosun}$\fi}
\def \hMsun {\ifmmode h^{-1}\,\mathrm M_{\astrosun}\else$h^{-1}\,\mathrm M_{\astrosun}$\fi}
\def \Gyr {\ifmmode\,\mathrm{Gyr}\else$\,$Gyr\fi}

\def \LCDM {\ifmmode\Lambda{\mathrm{CDM}}\else$\Lambda{\mathrm{CDM}}$\fi}
\def \sig8 {\ifmmode\sigma_8\else$\sigma_8$\fi} 
\def \OmegaM {\ifmmode\Omega_{\mathrm{m}}\else$\Omega_{\mathrm{m}}$\fi} 
\def \Omegab {\ifmmode\Omega_{\mathrm{b}}\else$\Omega_{\mathrm{b}}$\fi} 
\def \OmegaL {\ifmmode\Omega_{\mathrm{\Lambda}}\else$\Omega_{\mathrm{\Lambda}}$\fi} 
\def \Deltavir {\ifmmode\Delta_{\mathrm{vir}}\else$\Delta_{\mathrm{vir}}$\fi}
\def \rhocrit {\ifmmode\rho_{\mathrm{crit}}\else$\rho_{\mathrm{crit}}$\fi}
\def \rhou {\ifmmode\rho_{\mathrm u}\else $\rho_{\mathrm u}$\fi}
\def \zc {\ifmmode z_{\mathrm c}\else $z_{\mathrm c}$\fi}

\title[HELLO project] {HELLO project: High-$z$ Evolution of Large and Luminous Objects}

\author[
S.\ Waterval et al.]{Stefan Waterval$^{\href{https://orcid.org/0000-0002-5542-8624}{\hskip2pt\includegraphics[width=9pt]{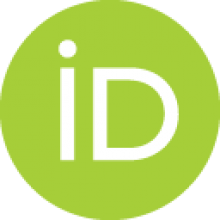}}{1,2}}$\thanks{E-mail: sw4445@nyu.edu},
Andrea V.\ Macciò$^{\href{https://orcid.org/0000-0002-8171-6507}{\hskip2pt\includegraphics[width=9pt]{figures/orcid-ID.png}}{1,2,3}}$, 
Tobias Buck$^{\href{https://orcid.org/0000-0003-2027-399X}{\hskip2pt\includegraphics[width=9pt]{figures/orcid-ID.png}}{4,5}}$,
Aura Obreja$^{\href{https://orcid.org/0000-0003-4196-8555}{\hskip2pt\includegraphics[width=9pt]{figures/orcid-ID.png}}{4,5}}$,
\newauthor{Changhyun Cho$^{\href{https://orcid.org/0000-0002-9879-1749}{\hskip2pt\includegraphics[width=9pt]{figures/orcid-ID.png}}{1,2}}$},
Zehao Jin$^{\href{https://orcid.org/0009-0000-2506-6645}{\hskip2pt\includegraphics[width=9pt]{figures/orcid-ID.png}}{1,2}}$,
Benjamin L.\ Davis$^{\href{https://orcid.org/0000-0002-4306-5950}{\hskip2pt\includegraphics[width=9pt]{figures/orcid-ID.png}}{1,2}}$,
Keri L.\ Dixon$^{\href{https://orcid.org/0000-0001-9311-1639}
{\hskip2pt\includegraphics[width=9pt]{figures/orcid-ID.png}}{1,2}}$,
\newauthor{and Xi Kang$^{\href{https://orcid.org/0000-0002-5458-4254}{\hskip2pt\includegraphics[width=9pt]{figures/orcid-ID.png}}{6,7}}$}
\\
$^{1}$New York University Abu Dhabi, PO Box 129188, Abu Dhabi, United Arab Emirates \\
$^2$Center for Astrophysics and Space Science (CASS), New York University Abu Dhabi\\  
$^3$Max-Planck-Institut f\"ur Astronomie, K\"onigstuhl 17, D-69117 Heidelberg, Germany\\
$^4$Interdisziplinäres Zentrum für Wissenschaftliches Rechnen, Universität Heidelberg,\\ Im Neuenheimer Feld 205, D-69120 Heidelberg, Germany\\
$^5$Zentrum für Astronomie, Institut für Theoretische Astrophysik, Universität Heidelberg,\\ Albert-Ueberle-Straße 2, D-69120 Heidelberg, Germany\\
$^6$Institute for Astronomy, Zhejiang University, Hangzhou 310027, China\\
$^7$Purple Mountain Observatory, 10 Yuan Hua Road, Nanjing 210034, China
}

\date{Accepted XXX. Received YYY; in original form ZZZ}

\pubyear{2024}

\begin{document}
\label{firstpage}
\pagerange{\pageref{firstpage}--\pageref{lastpage}}
\maketitle

\begin{abstract}
We present the High-$z$ Evolution of Large and Luminous Objects (HELLO) project, a set of $\sim\!30$ high-resolution cosmological simulations aimed to study Milky Way analogues ($M_\star\sim10^{10-11}$\,\Msun) at high redshift ($z\sim [2-4]$).
Based on the Numerical Investigation of a Hundred Astrophysical Objects (NIHAO),
HELLO features an updated scheme for chemical enrichment and the addition of local photoionization feedback.
Independently of redshift and mass, our galaxies exhibit a smooth progression along the star formation main sequence until $M_\star \sim\!10^{10.5}$, around which our sample at $z \sim 4$ remains mostly unperturbed while the most massive galaxies at $z \sim 2$ reach their peak star formation rate (SFR) and its subsequent decline, due to a mix of gas consumption and stellar feedback. While AGN feedback remains subdominant with respect to stellar feedback for energy deposition, its localised nature likely adds to the physical processes leading to declining SFRs. The phase in which a galaxy in our mass range can be found at a given redshift is set by its gas reservoir and assembly history.
Finally, our galaxies are in excellent agreement with various scaling relations observed with the \textit{Hubble Space Telescope} and the \textit{James Webb Space Telescope}, and hence can be used to provide the theoretical framework to interpret current and future observations from these facilities and shed light on the transition from star-forming to quiescent galaxies.

\end{abstract}

\begin{keywords}
quasars: supermassive black holes, galaxies: formation, galaxies: evolution, methods: numerical, methods: statistical
\end{keywords}

%%%%%%%%%%%%%%%%%%%%%%%%%%%%%%%%%%%%%%%%%%%%%%%%%%%
\section{Introduction}\label{sec:introduction}
%%%%%%%%%%%%%%%%%%%%%%%%%%%%%%%%%%%%%%%%%%%%%%%%%%%

%\citep[txt1][txt2]{citation}
% gives (txt1 Name et al. 2020, txt2)

%citet{citation}
%gives Name et al. (2020)

%citealt{citation}
%gives Name et al. 2020

As we enter an era marked by the deployment of more advanced astronomical instruments capable of observing galaxies nearer to the dawn of the Universe, such as the \textit{James Webb Space Telescope} \citep[\textit{JWST};][]{Gardner2006}, the Extremely Large Telescope \citep[ELT;][]{Neichel2018}{}{}, the Thirty Meter Telescope \citep[TMT;][]{Skidmore2015}{}{}, and others, establishing a solid theoretical foundation for galaxy formation and evolution at high redshift becomes increasingly crucial.
Star-forming galaxies (SFGs) from the early epochs are the precursors to today's elliptical and massive disk galaxies, including our own Milky Way (MW).
These early galaxies provide a unique window into the complex nature of these systems.
Consequently, it is vital for cosmological simulations to accurately replicate observed phenomena, thereby enhancing our comprehension of the evolutionary pathways of galaxies.

High-$z$ SFGs consistently show significantly higher star-formation activity than local analogues \citep{Schiminovich2005, LeFloch2005} with star-formation rates (SFRs) in the tens and hundreds of \Msun\,yr$^{-1}$ \citep[see, e.g.,][]{Gruppioni2013}.
Additionally, both local SFGs and their higher-$z$ counterparts exhibit a tight correlation between SFR and stellar mass (SFR--$\Mstar$ relation), with increasing normalisation at later cosmic times \citep{Brinchmann2004, Noeske2007, Elbaz2007, Daddi2007}.
This SFR--$M_\star$ relation, also referred to as the star-formation main sequence (SFMS), is commonly parameterised as a power law, such that $\mathrm{SFR} \propto M_\star^{\alpha}$, with slope $\alpha$ about unity for low-mass galaxies ($M_\star \lesssim 10^{10}$\,\Msun).
The presence of a shallower slope above a turn-over mass is still contested, with, e.g., \citet{Whitaker2014}, \citet{Lee2015}, \citet{Schreiber2015}, \citet{Tomczak2016}, and \citet{Leja2022} finding that star formation slows down at the high-mass end, while \citet{Speagle2014} and \citet{Pearson2018} find a SFMS consistent with a single power law at fixed $z$.
Nevertheless, such high SFRs measured at early epochs need to eventually be suppressed in order to produce local galaxies with stellar masses consistent with observations.
This can be achieved either by removing gas, heating it, hindering it from cooling, or keeping it from collapsing altogether.
The physical origins of the quenching process are still debated and different scenarios have been proposed.

First, active galactic nuclei (AGNs) in the centres of massive galaxies are thought to be crucial in removing gas and inhibiting the cooling processes via injection of energy and momentum in their surroundings, thus leading to the quenching of star formation observed in elliptical galaxies \citep[][]{McNamara2007, Hardcastle2013, Ellison2016, Leslie2016, Comerford2020}, however, cf., e.g., \citet{Juneau2013, Bernhard2016}, and \citet{Dahmer-Hahn2022}, for opposite claims.
In fact, AGN feedback is required in most simulations to lower star formation and recover numerous key observables of massive galaxies \citep[][]{Valageas1999, Vogelsberger2014, Crain2015, Costa2018, Blank2019, Zinger2020}.
Second, \citet{Martig2009} have proposed a method of `morphological quenching', whereby the buildup of a central bulge with a sufficiently deep potential well stabilises the gaseous disk around it, which in turn hampers the ability for gas to collapse into bound clumps.
Finally, a type of `halo quenching', suggests that infalling gas in haloes above a mass $\sim$$10^{12}$\,\Msun~can be prevented from cooling efficiently via shock-heating \citep[][]{Birnboim2003, Keres2005, Dekel2006}.

Beyond the SFMS, a collection of other well-studied scaling relations are seemingly already in place at early times.
\citet{vanderWel2014} find that the effective radius ($\Reff$) of early-type galaxies and SFGs alike, exhibits a tight scaling with stellar mass at all redshifts $0 < z < 3$.
The two populations' size--mass relations differ, however, with SFGs being larger at all masses and manifesting a relatively constant and flatter slope of $\Reff \propto \Mstar^{0.22}$ across all redshifts probed.
While the size--mass relation holds for the entire population, individual galaxies could actually have evolved along steeper tracks, from $\Reff \propto \Mstar^{0.27-0.3}$ for the progenitors of today's MW-mass systems \citep[][]{VanDokkum2013, VanDokkum2015} up to $\Reff \propto \Mstar^{2}$ for the most massive galaxies \citep[][]{Patel2013}.
In addition, galaxies at a fixed mass become smaller with increasing redshift \citep[][]{Trujillo2007, Buitrago2008, vanderWel2012}{}{}, a trend recently confirmed with \textit{JWST} observations \citep[][]{Ormerod2024}.

% Perhaps the most discussed findings so far from the \textit{JWST} pertain to the unexplained prevalence of undermassive galaxies, relative to the comparatively overmassive quasars that they host.
% For example, \citet{Stone:2023} find five quasars at $z\sim5$--7, with black hole masses ($M_\bullet$) ranging from $\log(M_\bullet/\Msun)=8.8$ to 9.8, residing in host galaxies with masses only as high as $\log(M_\star/\Msun)=9.7$ to 10.8.
% This places the claimed quasars well above the local $M_\bullet$--\Mstar~scaling relation.
% Specifically, the local $M_\bullet$--\Mstar~relation \citep{Davis:2018,Sahu:2019,Graham:2023} predicts $M_\bullet/\Mstar$ ratios of $\sim$$10^{-4}$--$10^{-3}$ (depending on morphology) for this mass range of galaxies, whereas \citet{Stone:2023} apparently find a median $M_\bullet/\Mstar$ ratio of $\sim$$10^{-1}$ at $z\sim6$.
% This emerging enigma from analyses of \textit{JWST} observations sets a clear precedence for numerical galaxy simulations to conduct pivotal checks of correspondence with observed galaxies to reproduce realistic galaxies that match observed scaling relations \citep[e.g.,][]{Sahu:2020,Sahu:2022}.

Another significant metric, the stellar surface density, either measured within 1\,kpc ($\Sigma_1$) or within \Reff\ ($\Sigmae$), is posited to be crucial in deciphering the mechanisms and timing of the transition of SFGs to the quiescent population, offering insights into their evolutionary trajectories.
The two variants have been shown to be tightly correlated with stellar mass and can help understand the quenching mechanisms in galaxies \citep[][]{Cheung2012, Fang2013, Barro2017}.
Indeed, numerous studies propose that quenching is preceded by a significant central density growth \citep[][]{Schiminovich2007, Bell2008, Lang2014, Whitaker2017}.
The buildup of such a bulge, however, is merely correlated with the quenching process and \textit{probably} not a causal physical source. Cosmological simulations suggest that high-$z$ galaxies experience a phase of wet compaction \citep[][]{Dekel2014}{}{} followed by a peak in star formation and subsequent inside-out quenching after depletion of the available gas \citep[][]{Zolotov2015, Ceverino2015, Tacchella2015, Tacchella2016, Lapiner2023}{}{}.

In this paper, we present the High-$z$ Evolution of Large and Luminous Objects (HELLO) project stemming from the same cosmological hydrodynamical code for galaxy formation as the Numerical Investigation of a Hundred Astrophysical Objects \citep[NIHAO;][]{Wang2015}{}{}.
Whereas NIHAO consists of a large statistical sample of high-resolution zoom-in simulations of local galaxies covering several orders of magnitude in mass, HELLO aims to study their massive ($\Mstar \sim 10^{10-11}\,\Msun$) counterparts at $z \sim$ [2--4] with roughly twice the resolution and includes $\sim$30 objects at the time of writing.
The NIHAO simulations have been remarkably successful in replicating a range of galaxy properties.
These include the stellar-to-halo mass relation (SHMR) as demonstrated by \citet{Wang2015}, the correlation between disk gas mass and disk size \citep[][]{Maccio2016}, and the Tully-Fisher relation as per \citet{Dutton2017}.

Additionally, the NIHAO simulations accurately represented the satellite mass function of the MW and M31, as evidenced in \citet{Buck2019}.
Finally, extensive works have studied the effects of AGN feedback on dark and baryonic matter in NIHAO \citep[e.g.,][]{Maccio2020, Blank2021, Waterval2022}.
We introduce our new set of HELLO galaxies by comparing them with a few key scaling relations observed at high-$z$, in particular the SHMR, the SFMS, the size--mass relation, and the $\Sigma$--$\Mstar$ relation.
Finally, we attempt to provide an explanation to the causes behind the declining SFRs seen in some of our galaxies by analysing the content, evolution, and distribution of gas, as well as the central black hole (BH) growth and corresponding energy feedback released.

This paper is organized as follows: in \S\ref{sec:simulations}, we specify the aspects of our simulations, including an accounting of our initial conditions (\S\ref{subsec:initial_conditions}), star formation and stellar feedback (\S\ref{subsec:star_formation}), chemistry (\S\ref{subsec:chemistry}), local photoionisation feedback (\S\ref{subsec:lpf}), and black hole growth and feedback (\S\ref{subsec:bh_implementation}); \S\ref{sec:sfh} is where we detail the star-formation rates and histories of galaxies in our simulations; we remark on the various scaling relations observed from our ensemble of simulated galaxies in \S\ref{sec:scaling_relations}, specifically, the stellar-to-halo mass relation (\S\ref{subsec:shmr}), the main sequence of star formation (\S\ref{subsec:sfr_mstar}), size-mass relation (\S\ref{subsec:size_mass_relation}), and surface density scaling relations (\S\ref{subsec:surf_den}); then in \S\ref{sec:gas_and_feedback}, we address the gas availability via a discussion of gas temperature and density (\S\ref{subsec:gas_temp_dens}), and subsequently AGN feedback (\S\ref{subsec:agn_feedback}) present in our simulations; and finally, we provide a concise summary and concluding remarks in \S\ref{sec:discussion}.

\begin{figure*}
\centering
\includegraphics[width=\textwidth]{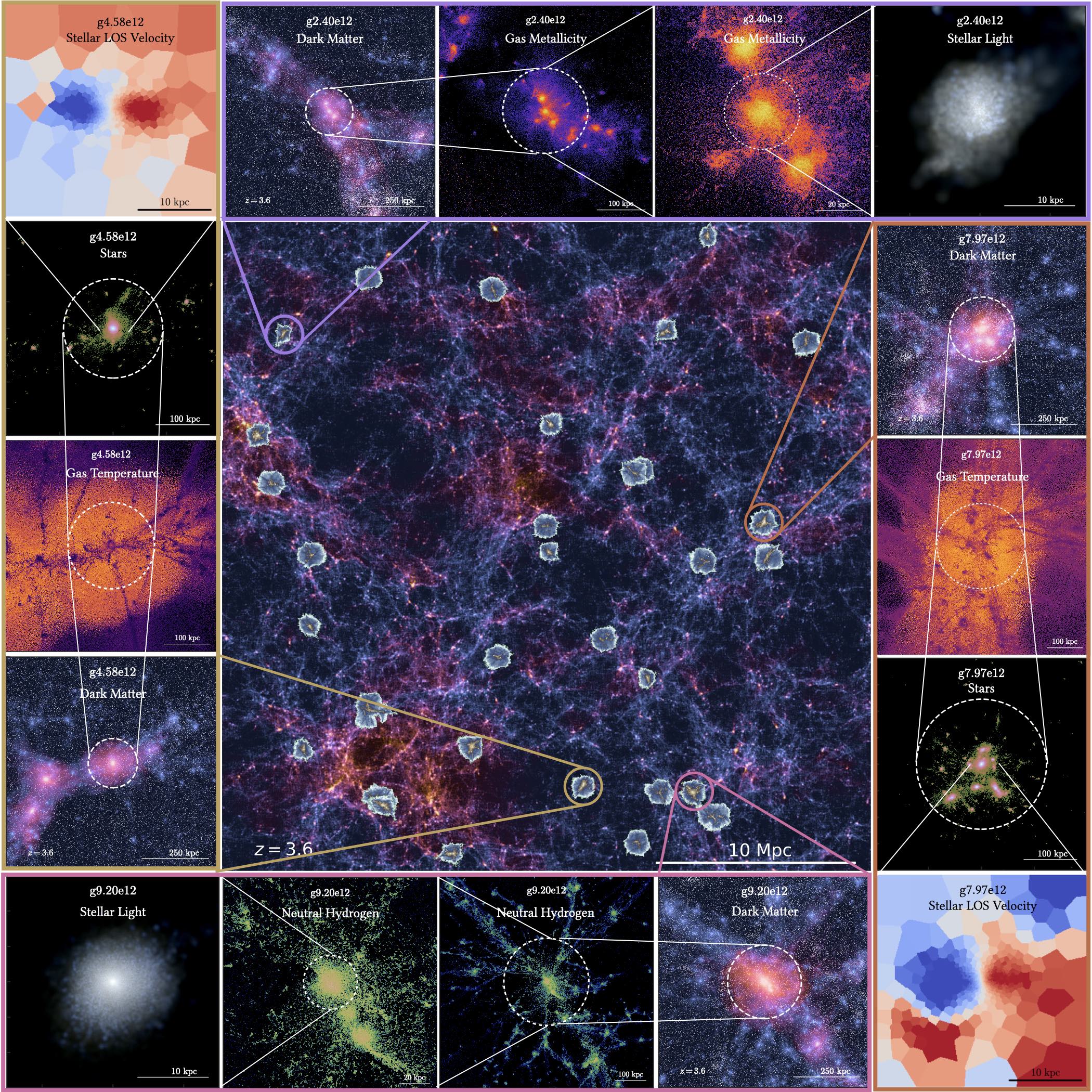}
\caption{
Highlights of the physical and numerical complexity of the `High-$z$ Evolution of Large and Luminous Objects' (HELLO) project.
The central panel shows a rendering of the dark matter distribution with the 32 zoom regions overlaid on top.
Four representative zoom regions are circled and the smaller panels surrounding the central panel show continuous zooms onto the central galaxy, imaging various physical quantities of the simulations, thereby highlighting the large dynamical range and physical complexity of the simulation suite.
}
\label{fig:collage}
\end{figure*}

%%%%%%%%%%%%%%%%%%%%%%%%%%%%%%%%%%%%%%%%%%%%%%%%%%%
\section{Simulations}\label{sec:simulations}
%%%%%%%%%%%%%%%%%%%%%%%%%%%%%%%%%%%%%%%%%%%%%%%%%%%
HELLO is a set of cosmological zoom-in hydrodynamical simulations at $z \sim [2, 4]$ and builds on NIHAO by using the same methods for generating initial conditions, star formation and stellar feedback, and black hole growth and feedback.
New implementations concern chemical element tracking and introduction of local photoionisation feedback (LPF).
For completeness, we summarise every part in the coming subsections.
The cosmological model employed is based on a flat $\Lambda$CDM framework, drawing on parameters from \citet{Planck2014}, and using a \textit{Hubble} constant ($H_0$) of 67.1\,$\kms\,\Mpc^{-1}$.
The densities associated with matter, dark energy, radiation, and baryons are given by \{$\Omega_\mathrm{m}$, $\Omega_\Lambda$, $\Omega_\mathrm{r}$, $\Omega_\mathrm{b}$\} = \{0.3175, 0.6824, 0.00008, 0.0490\}.
Additionally, the power spectrum's normalisation is set to $\sigma_8$ = 0.8344 and its slope to $n$ = 0.9624.

The HELLO project presently consists of two sets of 17 and 15 galaxies whose haloes contain about $N_\mathrm{part} \sim 10^6$ particles within their virial radii at the final redshifts of 2.0 (age $\sim$ 3.3\,Gyr) and 3.6 (age $\sim$ 1.7\,Gyr), respectively (see Fig.~\ref{fig:collage} for a visual overview of the simulation volume).
For the rest of this work, these two sets will be denominated `HELLOz2.0' and `HELLOz3.6.'
Haloes in HELLO are identified using the Amiga Halo Finder \citep[\ahf;][]{Gill2004, Knollmann2009}{}{}.
The virial radius is defined as the radius within which the halo contains an average density 200 times the critical density $\rho_\mathrm{c}(z)$ at redshift $z$ and is denoted as $R_{200}$.

The total mass enclosed in a sphere of radius \Rtwohun\ is thus expressed as \Mtwohun\ and both are related by:
\begin{equation}
    \Mtwohun = \frac{4}{3}\,\pi\,\Rtwohun^3\,200\,\rho_{\mathrm{c}}.
\end{equation}
In the remainder of the paper, we use \Mtwohun\ and \Mhalo\ interchangeably.
The simulations' final halo masses range from $\sim$(1 to $9)\,\times 10^{12}\,\Msun$.
HELLO thus aims at studying massive MW-like galaxies around cosmic noon \citep[][]{MadauDickinson2014}{}{}. We emphasise that our galaxies are not progenitors of today's MW-like galaxies, but likely of more massive ellipticals.

Each halo is evolved hydrodynamically with the \textsc{\small GASOLINE2} code \citep[][]{Wadsley2017}, while a dark matter only (DMO) counterpart is run with \textsc{\small PKDGRAV2} \citep[][]{Stadel2001, Stadel2013}.
Dark matter mass and gas mass resolution in the hydro versions amount to $3.4 \times 10^6$\,\Msun\ and $2.1 \times 10^5$\,\Msun, respectively, with corresponding physical softening lengths of $\epsilon_\mathrm{DM} = 193$ (127)\,pc and $\epsilon_\mathrm{gas} = 83$ (54)\,pc at $z=2$ (3.6).
Some of the principal quantities in our simulations are summarised in Table~\ref{tab:ic_params}.
For an exhaustive list of all our simulation parameters at their respective final redshifts, see Appendix~\ref{app:hello_quantities}.
\begin{table*}
    \centering
    \begin{tabular}{lrrrrrr}
        \hline
        Name & $z_{\mathrm{final}}$ & $M_\mathrm{halo}$\,[$10^{12}$\,\Msun] & $M_\mathrm{gas}$\,[$10^{10}$\,\Msun] & $M_\star$\,[$10^{10}$\,\Msun] & $R_{200}$\,[kpc] & $N_\mathrm{part}$\\
        \hline
        g3.08e12 & 2.0 & 2.84 & 7.31 & 10.1 & 142 & 1,943,368\\
        g3.09e12 & 2.0 & 2.94 & 7.26 & 11.3 & 144 & 2,058,093\\
        g3.00e12 & 2.0 & 2.43 & 8.33 & 5.91 & 135 & 1,501,948\\
        g3.20e12 & 2.0 & 2.72 & 9.27 & 5.69 & 140 & 1,693,806\\
        g2.75e12 & 2.0 & 2.45 & 8.77 & 9.74 & 135 & 1,778,286\\
        g3.03e12 & 2.0 & 1.67 & 5.69 & 7.55 & 119 & 1,250,411\\
        g3.01e12 & 2.0 & 2.67 & 7.41 & 7.67 & 139 & 1,729,071\\
        g2.29e12 & 2.0 & 1.81 & 7.71 & 7.53 & 122 & 1,333,337\\
        g3.35e12 & 2.0 & 1.72 & 6.69 & 2.37 & 120 & 964,316\\
        g3.31e12 & 2.0 & 2.32 & 8.21 & 5.27 & 133 & 1,449,371\\
        g3.25e12 & 2.0 & 2.65 & 8.11 & 4.09 & 139 & 1,643,064\\
        g3.38e12 & 2.0 & 3.08 & 9.86 & 8.09 & 146 & 2,014,002\\
        g3.36e12 & 2.0 & 2.08 & 4.43 & 3.14 & 128 & 1,209,250\\
        g2.83e12 & 2.0 & 2.32 & 5.75 & 6.57 & 133 & 1,551,228\\
        g2.63e12 & 2.0 & 2.35 & 7.85 & 9.01 & 133 & 1,682,819\\
        g3.04e12 & 2.0 & 3.01 & 8.10 & 9.50 & 145 & 1,997,107\\
        g2.91e12 & 2.0 & 2.11 & 4.33 & 7.29 & 129 & 1,445,948\\
        \hline
        g2.47e12 & 3.6 & 2.21 & 8.16 & 3.39 & 87 & 1,259,900\\
        g2.40e12 & 3.6 & 2.29 & 5.16 & 0.62 & 88 & 1,231,536\\
        g2.69e12 & 3.6 & 2.54 & 11.00 & 4.53 & 91 & 1,493,797\\
        g3.76e12 & 3.6 & 2.58 & 12.30 & 3.45 & 102 & 1,998,769\\
        %g4.36e12 & 3.6 & 2.66 & 8.98 & 2.65 & 93 & 1,506,328\\
        g4.58e12 & 3.6 & 2.60 & 11.20 & 5.86 & 92 & 1,616,009\\
        g2.71e12 & 3.6 & 1.98 & 9.75 & 4.35 & 84 & 1,216,000\\
        g2.49e12 & 3.6 & 2.55 & 9.72 & 2.99 & 91 & 1,427,551\\
        g2.51e12 & 3.6 & 2.52 & 6.57 & 2.61 & 91 & 1,529,843\\
        g2.32e12 & 3.6 & 1.91 & 8.47 & 4.93 & 83 & 1,218,031\\
        g2.96e12 & 3.6 & 2.84 & 12.70 & 3.43 & 95 & 1,605,548\\
        g7.37e12 & 3.6 & 5.76 & 20.70 & 11.60 & 120 & 3,526,490\\
        g7.97e12 & 3.6 & 5.89 & 13.80 & 7.12 & 121 & 3,546,175\\
        g9.20e12 & 3.6 & 8.74 & 25.70 & 14.10 & 138 & 5,318,120\\
        %g9.85e12 & 3.6 & 9.10 & 29.40 & 21.70 & 140 & 5,793,519\\
        \hline
    \end{tabular}
    \caption{
    Final HELLO sample at $z=2.0$ and $z=3.6$ used in this paper.
    The columns are: galaxy name (a reference to the low-resolution halo mass at the final redshift); final redshift ($z_\mathrm{final}$); halo ($M_\mathrm{halo}$), gas ($M_\mathrm{gas}$), and stellar ($M_\star$) masses within 20~per~cent of the virial radius (\Rtwohun); and total number of particles in the halo ($N_\mathrm{part}$).
    }
    \label{tab:ic_params}
\end{table*}

Unless otherwise specified, galactic quantities such as, e.g., $\Mstar$, gas mass ($M_\mathrm{gas}$), $\Reff$, and SFR are computed using the corresponding particles within 20~per~cent of \Rtwohun\ to remain consistent with NIHAO simulations.
While small satellites may be present at these distances, we control the resulting stellar mass, effective radius, effective stellar surface density, and SFR so they do not vary by more than $\sim$10~per~cent with respect to using $0.1\,\Rtwohun$. The two galaxies that didn't pass these criteria were removed from the remaining calculations.
A more detailed discussion can be found in Appendix~\ref{app:satellite_contamination}.

\begin{figure*}
\centering
\includegraphics[clip=true, trim= 2mm 2mm 2mm 2mm, width=0.85\textwidth]{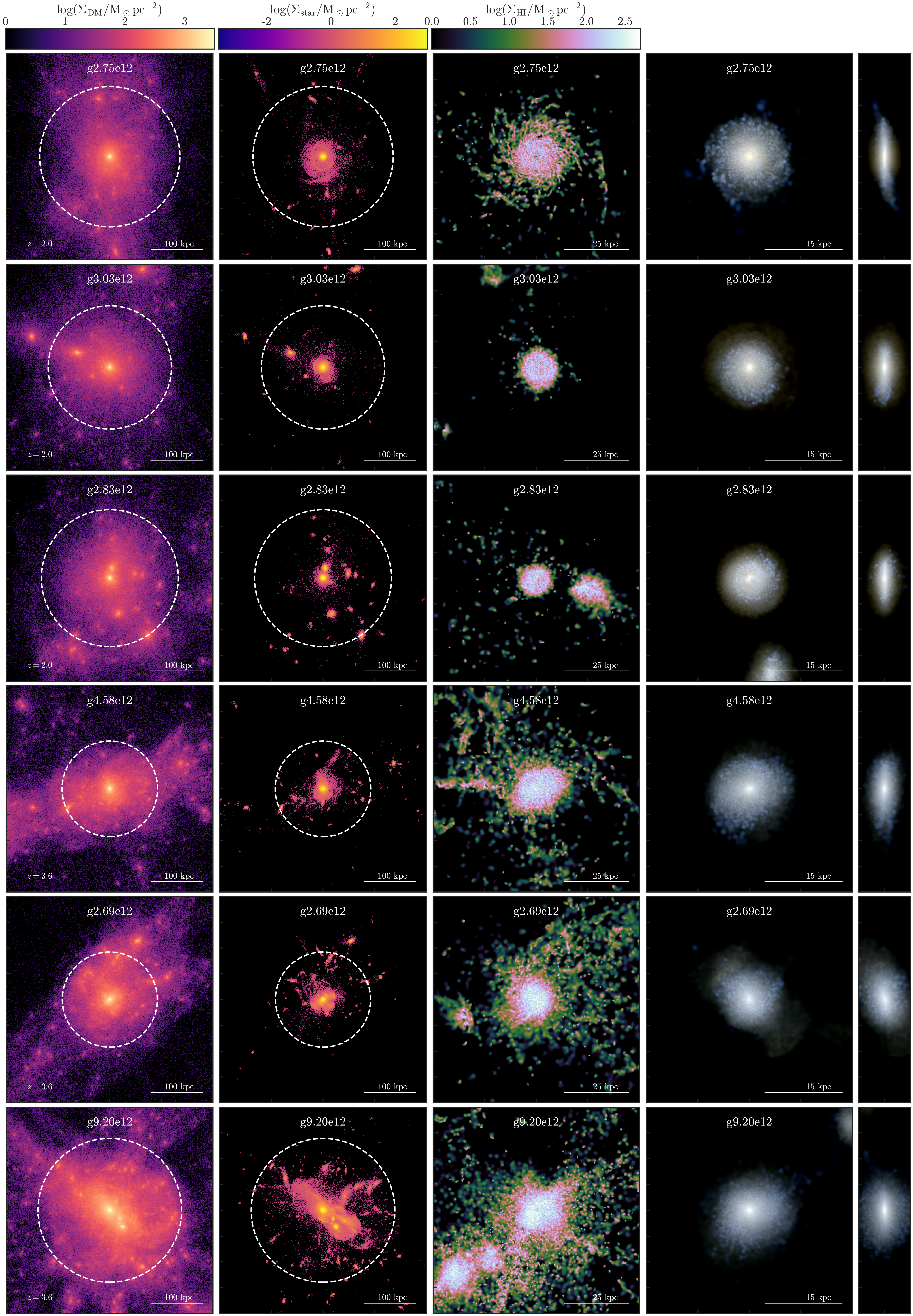}
\caption{Surface density maps viewed face-on as well as \textbf{dust-free} stellar face- and edge-on images in the wavelength bands $I$, $V$, and $U$ for six example galaxies (three at $z=2$ and three at $z=3.6$).
The surface density columns from left to right are: dark matter, stars, and \ion{H}{I} gas which is based on the fitting formulas of \citet{Rahmati2013}.
The white dashed circle indicates the virial radius ($\Rtwohun$).}
\label{fig:surf_den_maps}
\end{figure*}

We show in Fig.~\ref{fig:surf_den_maps}, a visual aper\c{c}u of six HELLO galaxies, three belonging to the HELLOz2.0 sample (top three rows) and three to the HELLOz3.6 sample (bottom three rows).
The first three columns from left to right are a face-on view of dark matter, stars, and \ion{H}{I} gas, respectively.
The last two columns show an image rendering in face- and edge-on views of stars in the wavelength bands $I$, $V$, and $U$.
The white dashed circle represents the virial radius, and the simulation name, as well as a distance scale is displayed in each panel.
Our simulations exhibit a variety of morphologies, from relatively thin disks to more compact and spherical objects, as well as different local environments with some galaxies like g2.75e12, g3.03e12, and g4.58e12 fairly isolated, while g2.83e12 and g9.20e12 are about to merge with surrounding satellites.

%%%%%%%%%%%%%%%%%%%%%%%%%%%%%%%%%%%%%%%%%%%%%%%%%%%
\subsection{Initial conditions}\label{subsec:initial_conditions}
%%%%%%%%%%%%%%%%%%%%%%%%%%%%%%%%%%%%%%%%%%%%%%%%%%%
All haloes are picked from a cosmological DMO volume simulation of size $100\,\hMpc^3$ at $z=0$ and containing $400^3$ particles with constant mass resolution, created with \textsc{\small PKDGRAV2}.
The motivation for these particular values is to allow for a sufficient number of virialised haloes of masses $M_{\mathrm{halo}} \sim 10^{12}$\,\Msun\ at the target redshifts, which can then be `zoomed-in' and evolved hydrodynamically.
The haloes are selected to be isolated, i.e., there are no other main structures more massive than 20~per~cent of the selected halo mass within $2\,\Rtwohun$. This selection process, however, does not consider the internal structure of the halo itself and does not preclude merger events at the galaxy level. For this reason, we added the aforementioned additional selection criterion on the final zoom simulations.
The zoom-in initial conditions are generated using an adapted version of the \textsc{\small GRAFIC2} software \citep[][]{Bertschinger2001}.\footnote{
We refer the reader to \citet{Penzo2014} for the specifics about these modifications.
}

The refinement level is set such as to keep a constant ratio between the dark matter (DM) force softening ($\epsilon_\mathrm{DM}$) and \Rtwohun, resolving the DM mass profile down to $\lesssim$$0.01\,\Rtwohun$.
We adopt here a DM softening length of 1/80 the inter-particle distance in the highest resolution region and the gas gravitational softening is calculated following $\epsilon_\mathrm{gas} = \epsilon_\mathrm{DM}\,\sqrt{\Omega_\mathrm{b}/\Omega_\mathrm{DM}}$, where $\Omega_\mathrm{DM} = \Omega_\mathrm{m} - \Omega_\mathrm{b}$.
The choice of 1/80 is motivated by the fact that it still satisfies the lower threshold set by the virial radius and the number of particles within \Rtwohun\ of $\epsilon > \Rtwohun/\sqrt{N_{200}}$, as suggested by \citet{power2003}, while better resolving the dynamical interactions than NIHAO (where the softening fraction is 1/40).
Moreover, recent work \citep[][]{Zhang2019}{}{} showed that softening fractions of 1/80 and 1/100 converge to smaller radii as compared to the optimal softening length proposed by \citet{power2003}.

%%%%%%%%%%%%%%%%%%%%%%%%%%%%%%%%%%%%%%%%%%%%%%%%%%%
\subsection{Star formation and stellar feedback}\label{subsec:star_formation}
%%%%%%%%%%%%%%%%%%%%%%%%%%%%%%%%%%%%%%%%%%%%%%%%%%%
The formation of stars follows the Kennicutt-Schmidt law \citep[][]{Schmidt1959, Kennicutt1998}{}{} and is implemented as described in \citet{Stinson2006, Stinson2013}.
First, to be eligible to form stars, gas particles have to satisfy both a temperature ($T < 15000$\,K) and density ($n > 10$\,cm$^{-3}$) threshold. The theoretical maximum gas density that we are able to resolve is $n_{\mathrm{max}} \sim 80$\,cm$^{-3}$, according to:
\begin{equation}
    n_{\mathrm{max}} \approx N_{\mathrm{smooth}}\, \frac{m_{\mathrm{gas}}}{\epsilon_{\mathrm{gas}}^3},
\end{equation}
where $N_{\mathrm{smooth}} = 50$ is the number of neighbouring particles within the smoothing kernel and $m_{\mathrm{gas}}$ is the gas particle initial mass.
In practice, however, gas can reach higher densities with a maximum value determined by the minimal smoothing length, $h_\mathrm{min} = 0.25\,\epsilon_\mathrm{gas}$.
In spite of halving the softening factor, we maintain the NIHAO threshold density of 10\,cm$^{-3}$, since higher thresholds have been shown to underproduce stars relative to abundance matching at $z=0$ for haloes with $\Mtwohun > 10^{11}\,\Msun$ \citep[][]{Dutton2020}.\footnote{
\citet{Dutton2020} do not use simulations with AGN feedback, which could further exacerbate the issue.
}
Gas particles satisfying the aforementioned criteria are allowed to convert to stars following:
\begin{equation}
    \frac{\Delta M_\star}{\Delta t} = c_\star\,\frac{M_{\mathrm{gas}}}{t_{\mathrm{dyn}}},
\end{equation}
where the mass of stars ($\Delta M_\star$) formed between each timestep ($\Delta t$) is a fraction $c_\star = 0.1$ of the theoretical SFR expressed as the eligible gas mass ($M_{\mathrm{gas}}$) that could form stars during one dynamical time ($t_{\mathrm{dyn}}$).

Stellar feedback is modeled in three different ways, depending on the epoch and stellar mass.
First, feedback energy from young bright stars is released in their vicinity during the first 4\,Myr after a star particle is born and is denoted as early stellar feedback (ESF).
The method adopts the model from \citet{Stinson2013} and is implemented as a fraction $\epsilon_\mathrm{ESF} = 0.13$ of the stellar luminosity being ejected as isotropic thermal energy into the surrounding gas.
The energy released by these stars typically represents $2 \times 10^{50}$\,erg\,$\Msun^{-1}$ over the first 4\,Myr after the creation of the stellar particle, and gas radiative cooling is left on during this epoch.

After the initial phase, supernovae (SNe) are initiated for stars with initial masses between 8 and 40\,\Msun\ within the stellar population.
The actual number of SNe is calculated from the stellar initial mass function (IMF), energy, mass, and metals ejected around the regions where the stars formed.
In practice, the energy released is $10^{51}$\,erg per SN and feedback from SNe is modelled according to the blast wave formalism described in \citet{Stinson2006}.
Affected gas particles see their radiative cooling turned off.
Finally, stars forming with masses below 8\,\Msun~constantly eject mass and metals as stellar winds.
% At each timestep, the mass range of stars that are supposed to die is computed and the corresponding returned mass fraction is determined \citep{Stinson2006}.

%%%%%%%%%%%%%%%%%%%%%%%%%%%%%%%%%%%%%%%%%%%%%%%%%%%
\subsection{Chemistry}\label{subsec:chemistry}
%%%%%%%%%%%%%%%%%%%%%%%%%%%%%%%%%%%%%%%%%%%%%%%%%%%
Star particles in cosmological simulations do not resolve individual stars, but are tracer particles representing a population of stars, referred to as a simple stellar population (SSP), that is described by a single age ($\tau_\mathrm{star}$), a metallicity ($Z_\mathrm{star}$), and an IMF specifying the number of stars in a given mass bin.
In our simulations, we chose a \citet{Chabrier2003} IMF.
We do not adopt a particular stellar evolution model, instead use the chemical evolution code \textsc{\small Chempy} \citep[][]{Rybizki2017} in order to synthesise the final stellar evolution model and the corresponding yield tables to be used for each run.
Hence, \textsc{\small Chempy} calculates the time evolution of the chemical yields for an SSP.
The stellar nucleosynthetic yields include the relative mass return fractions from three channels: type Ia supernova (SN\,Ia), core-collapse supernova (SN\,II), and asymptotic giant branch (AGB) stars.

Elemental feedback from these three channels is implemented as described in \citet{Buck2021}.
We tabulate the time-resolved, mass-dependent element release of a single stellar population as a function of initial metallicity in a grid of 50 metallicity bins, logarithmically-spaced between $10^{-5}$--0.05 in metallicity.
Each metal bin is resolved by 100 time bins, logarithmically-spaced in time and ranging from 0 to 13.8\,Gyr (the age of the Universe in our adopted cosmographic parameters).
Our fiducial combination of yield tables uses SN\,Ia yields from \citet{Seitenzahl2013}, SN\,II yields from \citet{Chieffi2004}, and AGB star yields from \citet{Karakas2016}.
All our simulations track the evolution of the 10 most abundant elements by default (H, He, O, C, Ne, Fe, N, Si, Mg, and S) while any other element present in the yield tables can additionally be tracked.
For HELLO galaxies, we also track the six following elements: Na, Al, Ca, Ti, Sc, and V.

%%%%%%%%%%%%%%%%%%%%%%%%%%%%%%%%%%%%%%%%%%%%%%%%%%%
\subsection{Local photoionisation feedback}\label{subsec:lpf}
%%%%%%%%%%%%%%%%%%%%%%%%%%%%%%%%%%%%%%%%%%%%%%%%%%%
High-energy ionisation photons from extragalactic sources are an important part of galaxy formation models and can provide a negative feedback affecting gas inflow and cooling \citep[][]{Haardt1996}.
While such external ultraviolet background (UVB) radiation can be approximated as isotropic on large cosmological scales, the local high-energy photon sources (stars, hot gas, or BHs) within the galaxy are distributed highly anisotropically.
Photoionisation and photoheating from three types of local radiation sources is implemented on top of the \citet{FaucherGiguere:2009} UVB in an optically-thin approximation, as described in \citet{Obreja2019}.
The three source types are: young stars ($\tau_\mathrm{star} < 10$\,Myr), post-AGB stars ($\tau_\mathrm{star} > 200$\,Myr), and bremsstrahlung from hot gas in three temperature bins: $5.5 < \log(T/\mathrm{K}) < 6.5$, $6.5 < \log(T/\mathrm{K}) < 7.5$, and $7.5 < \log(T/\mathrm{K}) < 8.5$.

We directly use the \textsc{\small Cloudy} \citep{Ferland1998} photoionisation models with the combinations of sources computed by \citet{Kannan2016}.
As young stars can still be partially enshrouded in their parent molecular clouds, only a fixed 5~per~cent of emitted photons are assumed to escape the local environment \citep[][]{Kannan2014}.
The other types of sources are assumed to have an escape fraction $f_\mathrm{esc}=1$.
Dense gas shielding from the local radiation field is modeled by a simple density-dependent relation, in which all gas particles with densities $n > 0.1$\,cm$^{-3}$ receive an exponentially attenuated local radiation field.
We prefer an attenuated field over an abrupt cutoff in order to emulate a smoother transition between shielded and partially-shielded regions.

%%%%%%%%%%%%%%%%%%%%%%%%%%%%%%%%%%%%%%%%%%%%%%%%%%%
\subsection{Black hole growth and feedback}\label{subsec:bh_implementation}
%%%%%%%%%%%%%%%%%%%%%%%%%%%%%%%%%%%%%%%%%%%%%%%%%%%
In HELLO, the implementation of BH and AGN feedback is performed according to \citet{Blank2019}.
Whenever the mass of a halo surpasses the threshold of $5 \times 10^{10}\,\Msun$, the gas particle with the lowest gravitational potential is converted into a seed BH of initial mass $M_{\mathrm{seed}} = 10^5\,\Msun$. 
The BH particle is then allowed to accrete and release feedback energy according to \citet{Springel2005b}.
The mass accretion is governed by Bondi accretion \citep[][]{Bondi1952}, scaled by a boost parameter, $\alpha = 70$:
\begin{equation}\label{eq:bondi}
    \dot{M}_{\mathrm{Bondi}} = \alpha\,\frac{4\pi\,G^2\,M_\bullet^2\,\rho_{\mathrm{gas}}}{(c_{\mathrm{s}}^2 + v_{\mathrm{gas}}^2)^{3/2}},
\end{equation}
with the BH mass $M_\bullet$, and $\rho_\mathrm{gas}$, $c_{\mathrm{s}}$, and $v_\mathrm{gas}$, the surrounding gas density, sound speed, and velocity, respectively.
The $\alpha$ parameter accounts for the finite resolution in simulations and its value is the optimal value found by \citet{Blank2019}.

The Eddington accretion rate \citep[][]{Eddington1921}:
\begin{equation}
    \dot{M}_{\mathrm{Edd}} = \frac{M_\bullet}{\epsilon_{\mathrm{r}}\,\tau_{\mathrm{S}}}
\label{eqn:Eddington}
\end{equation}
defines an upper bound for the accretion, as it is the rate at which the gravitational infall of gas is counter-balanced by the radiative pressure.
In Equation~\ref{eqn:Eddington}, the Salpeter timescale \citep[$\tau_{\mathrm{S}}$,][]{Salpeter:1964} and the radiative efficiency ($\epsilon_{\mathrm{r}}$) are set to 450\,Myr and 0.1, respectively.

During each timestep ($\Delta t$), the mass ($\dot{M}_\bullet\,\Delta t$) is accreted from the most gravitationally-bound gas particle, with $\dot{M}_\bullet \equiv \min\,\{\dot{M}_{\mathrm{Bondi}},\, \dot{M}_{\mathrm{Edd}}\}$.
If the gas particle's mass is less than 20~per~cent of its initial mass, the particle is removed from the simulation and its mass and momentum are distributed and weighted among the neighbours within the smoothing kernel.
The increase of $\Delta M_\bullet$ in BH mass is accompanied by an increase of its softening length to avoid too small steps due to large accelerations.
Finally, the thermal energy feedback resulting from accretion can be expressed as a fraction $\epsilon_{\mathrm f}= 0.05$ of the BH luminosity $L$:
\begin{equation}\label{eq:agn_feedback}
    \dot{E} = \epsilon_{\mathrm f}\,\dot{L} = \epsilon_{\mathrm f}\,\epsilon_{\mathrm r}\,\dot{M}_\bullet\,c^2.
\end{equation}
The thermal energy is distributed among the 50 nearest gas particles around the BH, weighted by the kernel function.

\begin{figure*}
\centering
\includegraphics[clip=true, trim= 6mm 6mm 6mm 6mm, width=\textwidth]{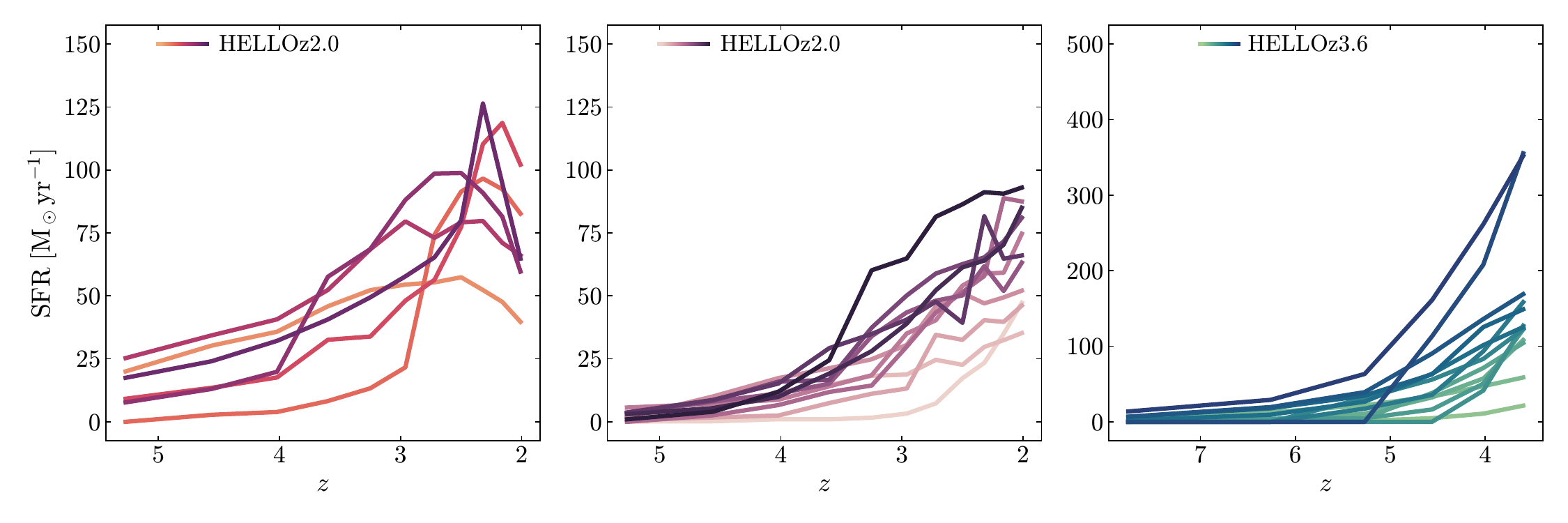}
\caption{
Individual SFHs of our simulations.
From left to right: HELLOz2.0 galaxies whose SFRs are declining towards their final redshift, HELLOz2.0 galaxies whose SFRs are still rising, and HELLOz3.6. 
The colour gradient maps the final stellar mass from lighter (lower mass) to darker (higher mass) shades.
}
\label{fig:sfh}
\end{figure*}

%%%%%%%%%%%%%%%%%%%%%%%%%%%%%%%%%%%%%%%%%%%%%%%%%%%
\section{Star-formation rate and history}\label{sec:sfh}
%%%%%%%%%%%%%%%%%%%%%%%%%%%%%%%%%%%%%%%%%%%%%%%%%%%
Our simulations confirm the higher SFRs observed in the early Universe.
At $z=2$, SFRs within the HELLO galaxies span from 35\,\Msun\,yr$^{-1}$ to 102\,\Msun\,yr$^{-1}$, with an average of 67\,\Msun\,yr$^{-1}$.
Earlier at $z=3.6$, the galaxies in our sample exhibit more star-forming activity, which translates to an increased average SFR of 152\,\Msun\,yr$^{-1}$, but also broader variations, with rates extending from 21\,\Msun\,yr$^{-1}$ to 355\,\Msun\,yr$^{-1}$.
These values are computed by summing up the formation mass of all stars located within $0.2\,\Rtwohun$ in the last 100\,Myr and averaging over the same timescale.
Our goal is to ensure the most appropriate comparison to observations as possible, since observations usually infer SFRs from a combination of infrared and ultraviolet luminosities, which are believed to trace the SFR of galaxies over the last\,100 Myr \citep[see, e.g.,][]{Speagle2014, MadauDickinson2014, ForsterSchreiber_2020}.
On top of our fiducial SFR, we also compute a more instantaneous SFR averaged over 10\,Myr, which we do not use in this paper, but include in Table~\ref{tab:appendix}, for completeness.

In Fig.~\ref{fig:sfh}, we plot the star formation history (SFH) of each galaxy in both our samples.
The colour gradients are sorted from light to dark by increasing stellar mass at the final redshift.
While all HELLOz3.6 galaxies exhibit rising SFRs up to their final redshift (right panel), the variety of SFH shapes is richer in the HELLOz2.0 sample. 
Indeed, some galaxies display an overall continuously-rising SFR (middle panel), but others seem to have reached their peak SFR and are on the decline (right panel).
A galaxy's SFR is considered to be declining if the SFR at $z=2$ is lower than in the last two snapshots.
Moreover, for the rest of this paper, we will assume that the galaxies with declining SFRs have begun their journey towards becoming quiescent, and we denote them as `post-peak,' as opposed to those that have not yet reached their peak SFR, designated as `pre-peak.'

\begin{figure}
\centering
\includegraphics[clip=true, trim= 4mm 4mm 4mm 4mm, width=\columnwidth]{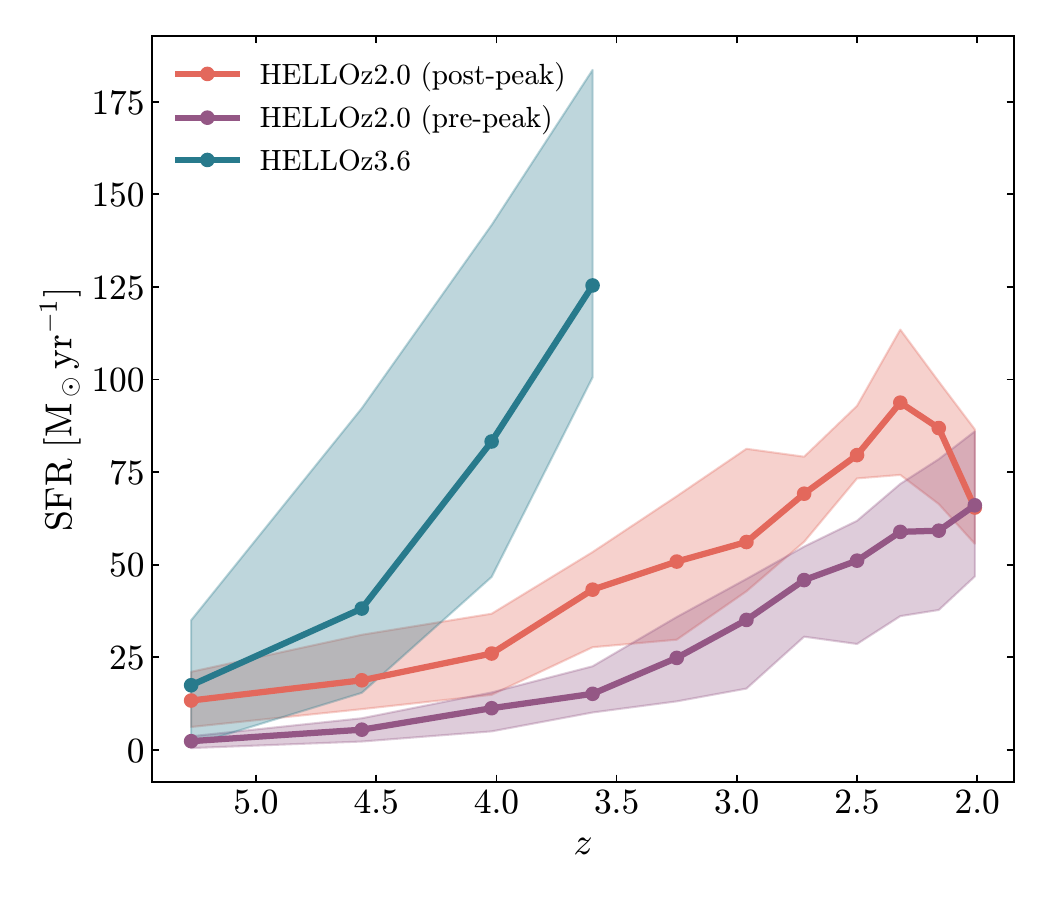}
\caption{
Median SFHs of HELLOz2.0 post-peak (orange), pre-peak (purple), and HELLOz3.6 (turquoise) galaxies.
The medians are calculated at each redshift from the curves in Fig.~\ref{fig:sfh}.
The shaded areas encompass the 16th and 84th percentiles.
}
\label{fig:sfh_median}
\end{figure}

We combine all SFHs in a single plot in Fig.~\ref{fig:sfh_median} by showing the median of each sample as thick curves (HELLOz2.0 pre- and post-peak: orange and purple; HELLOz3.6: turquoise), and the region encompassing the 16th to 84th percentiles as a shaded area.
Fig.~\ref{fig:sfh_median} gives a clear picture of the different regimes in which our simulations find themselves.
HELLOz3.6 galaxies are significantly more efficient at forming stars than their HELLOz2.0 counterparts, which comes at no surprise given that they reach similar end masses in roughly half the time.
On the other hand, both pre- and post-peak simulations feature a slower evolution until both samples reach a median SFR of $65\,\Msun\,\mathrm{yr}^{-1}$.
Post-peak galaxies, however, consistently have a higher median SFR, peaking at $93\,\Msun\,\mathrm{yr}^{-1}$ around $z=2.3$.
Importantly, the final stellar masses in the pre-peak sample range from $2.4 \times 10^{10}$ to $9.0 \times 10^{10}\,\Msun$, somewhat below the post-peak sample ranging from $7.3 \times 10^{10}$ to $1.1 \times 10^{11}\,\Msun$.
That is, post-peak galaxies are slightly more massive and thus, might be `further' in their evolution.

From Figs.~\ref{fig:sfh} and \ref{fig:sfh_median}, it is clear that none of our galaxies are quiescent, especially at $z=2$.
\citet{Muzzin2013}, using the UltraVISTA catalogue \citep[][]{McCracken2012}{}{}, found a fraction $0.25_{-0.12}^{+0.14}$ of quiescent galaxies at $2 \leq z < 2.5$, which for $\sim$15 galaxies should result in $\sim$2--6 quenched galaxies.
This could indicate some limitation in our AGN feedback model, which we briefly address at the end of \S\ref{subsec:agn_feedback}. Additionally, using cosmological zoom simulations from the MassiveFIRE \citep[][]{Feldmann2016}{}{} suite, \citet{Feldmann2017} found roughly 30~per~cent of their central galaxies to be quiescent at $z=2$. While they do not include AGN feedback, seven out of nine central galaxies identified as quiescent in their Fig.~2 have a halo mass larger than our most massive $z=2$ halo. Had we selected a larger range of halo masses at $z=2$ extending to $\log(M/\Msun) \sim [12.5-13]$, our sample would likely contain some quenched galaxies.

It is noteworthy that assessing the quiescent fraction within a selected population is far from trivial. Recent studies, using both simulations \citep[][]{Donnari2019} and observations \citep[][]{Leja2022, Neufeld2024}, have highlighted the variability and complexity in the identification and characterisation of quiescent galaxies. \citet{Donnari2019} demonstrated that the method of selection---whether through rest-frame colour or bimodality in the SFR--\Mstar\ plane---significantly impacts the fraction of quiescent galaxies (10--20~per~cent differences) at high masses. Similarly, \citet{Leja2022} found a $\sim $20~per~cent variation in the quiescent fraction obtained from four different methods using the same dataset. Finally, \citet{Neufeld2024}, using JWST data, identified sources that, despite being classified as quiescent in a typical UVJ diagram, exhibit significant Paschen-$\alpha$ emission, often extended in nature.

%%%%%%%%%%%%%%%%%%%%%%%%%%%%%%%%%%%%%%%%%%%%%%%%%%%
\section{Scaling relations}\label{sec:scaling_relations}
%%%%%%%%%%%%%%%%%%%%%%%%%%%%%%%%%%%%%%%%%%%%%%%%%%%

\begin{figure*}
\centering
\includegraphics[clip=true, trim= 4mm 4mm 4mm 4mm, width=\linewidth]{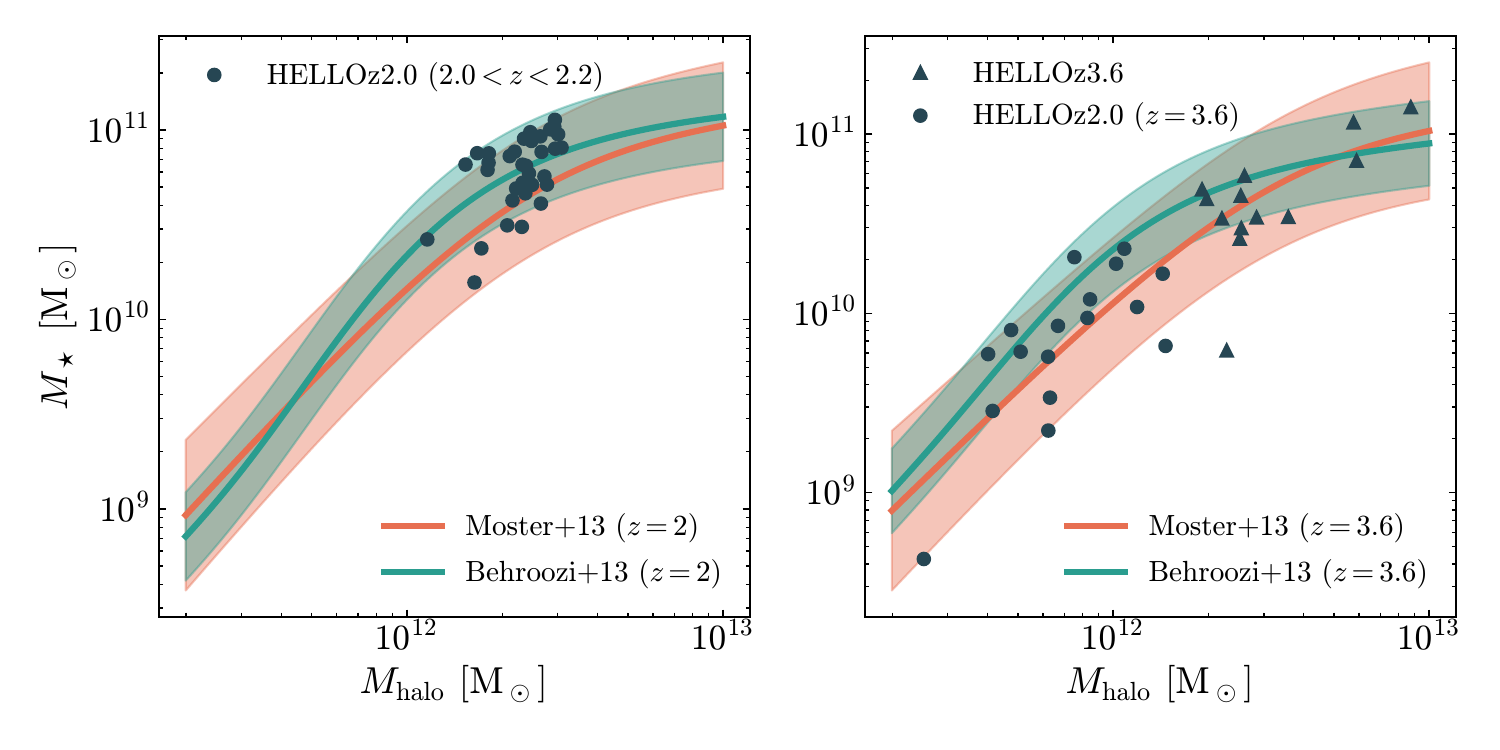}
\caption{
Stellar-to-halo mass relation for HELLO galaxies compared to abundance matching expectations from \citet{Moster2013} (orange curve) and \citet{Behroozi2013} (green curve) at $z=2$ (left panel) and $z=3.6$ (right panel).
The shaded area represents the $\pm1\,\sigma$ scatter around the respective relations.
In the left, we show HELLOz2.0 snapshots between redshifts 2.0 and 2.2, while on the right we show HELLOz3.6 final snapshots, as well as HELLOz2.0 progenitors at $z=3.6$.
}
\label{fig:shmr}
\end{figure*}

%%%%%%%%%%%%%%%%%%%%%%%%%%%%%%%%%%%%%%%%%%%%%%%%%%%
\subsection{Stellar-to-halo mass relation}\label{subsec:shmr}
%%%%%%%%%%%%%%%%%%%%%%%%%%%%%%%%%%%%%%%%%%%%%%%%%%%

We plot the SHMR of HELLO galaxies in Fig.~\ref{fig:shmr} and compare it against expectations from abundance matching \citep{Moster2013, Behroozi2013} at $z=2$ (left panel) and $z=3.6$ (right panel).
The shaded regions show the $\pm1\,\sigma$ scatter around the respective relations.
Black dots represent HELLOz2.0 galaxies between redshifts $2.0 < z < 2.2$ (left) and at $z=3.6$ (right).
Triangles in the right panel are the final snapshots of the HELLOz3.6 sample, while circles are HELLOz2.0 progenitors at $z=3.6$.

Our simulations agree well with both relations, as expected, since NIHAO galaxies have already been shown to match the $\Mstar$--$\Mtwohun$ relation well up to $z=4$ \citep{Wang2015, Buck2017, Blank2019}.
We experimented with a higher $n=80$\,cm$^{-3}$, but the resulting galaxies underproduced stars at $z=0$, an issue already covered in \S\ref{subsec:star_formation}.
Using such a value would require a lengthy recalibration of the feedback efficiencies, which is left to future studies.
For this work, although we found no significant discrepancies between fiducial galaxies run with $n=10$\,cm$^{-3}$ and $n=80$\,cm$^{-3}$ at high redshift, we required the SHMR to be in agreement at $z=0$.

%%%%%%%%%%%%%%%%%%%%%%%%%%%%%%%%%%%%%%%%%%%%%%%%%%%
\subsection{The main sequence of star formation}\label{subsec:sfr_mstar}
%%%%%%%%%%%%%%%%%%%%%%%%%%%%%%%%%%%%%%%%%%%%%%%%%%%
% Schreiber+15 takes UV+IR, i.e. average over 100 Myr
% Tomczak+16 takes UV+IR, i.e. average over 100 Myr

\begin{figure}
\centering
\includegraphics[clip=true, trim= 6mm 6mm 6mm 6mm, width=\columnwidth]{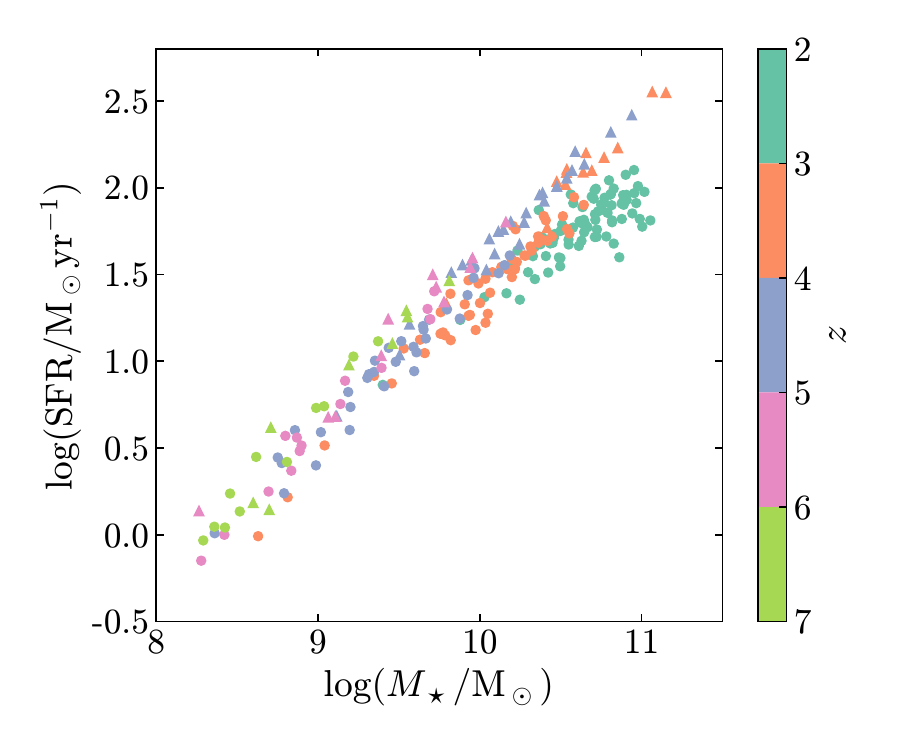}
\caption{
Stellar mass versus SFR of all our simulation snapshots, colour-coded in redshift bins from $z=2$ to $z=7$.
Galaxies belonging to the HELLOz2.0 sample are marked as circles, while those belonging to HELLOz3.6 are marked as triangles.
}
\label{fig:sfr_mstar_all_snaps}
\end{figure}

We begin by plotting the SFR--$M_\star$ relation for our galaxies in Fig.~\ref{fig:sfr_mstar_all_snaps}.
The figure contains all snapshots of all galaxies from both HELLOz2.0 (circles) and HELLOz3.6 (triangles) samples, colour-coded in redshift bins from $z=7$ down to $z=2$.
Each point represents the SFR from stars that formed in the last 100\,Myr of each snapshot, plotted against the respective stellar mass of each galaxy.
We find that HELLO galaxies are relatively tightly correlated across the entire sample.

The general consensus among observations is that the scatter around the main sequence is roughly constant at 0.2--0.3\,dex across the entire mass range, as well as redshifts probed \citep[][]{Speagle2014}, and simulations do agree overall with this picture \citep[e.g.,][]{Kannan2014, Torrey2014, Sparre2015}.
Qualitatively, Fig.~\ref{fig:sfr_mstar_all_snaps} exhibits low scatter at all masses and in each redshift bin, as well.
As previously mentioned, unanimity regarding the existence of a turnover mass and flattening of the star-forming main sequence (SFMS) has yet to be found.
However, when apparent, it is encountered at lower redshifts and tends to vanish above $z \sim 2$.

HELLO galaxies exhibit a similar behaviour, where the relation between $z = 3$--5 (orange and blue points) does not show any sign of a shallower slope. 
The same cannot be said for the turquoise points in redshift bin 2--3 that seem to flatten above $\log(M_\star/\Msun) \sim 10.5$.
Finally, for each redshift bin containing the progenitors of the galaxies in the next one (from green to turquoise), our plot shows that HELLO simulations evolve smoothly along the SFMS across cosmic times.
Interestingly, this coevolution between HELLOz3.6 and HELLOz2.0 seems to break around $\log(M_\star/\Msun) \sim 10$, above which HELLOz2.0 galaxies show signs of weakening star formation growth.

\begin{figure}
\centering
\includegraphics[clip=true, trim= 3mm 3mm 3mm 3mm, width=\columnwidth]{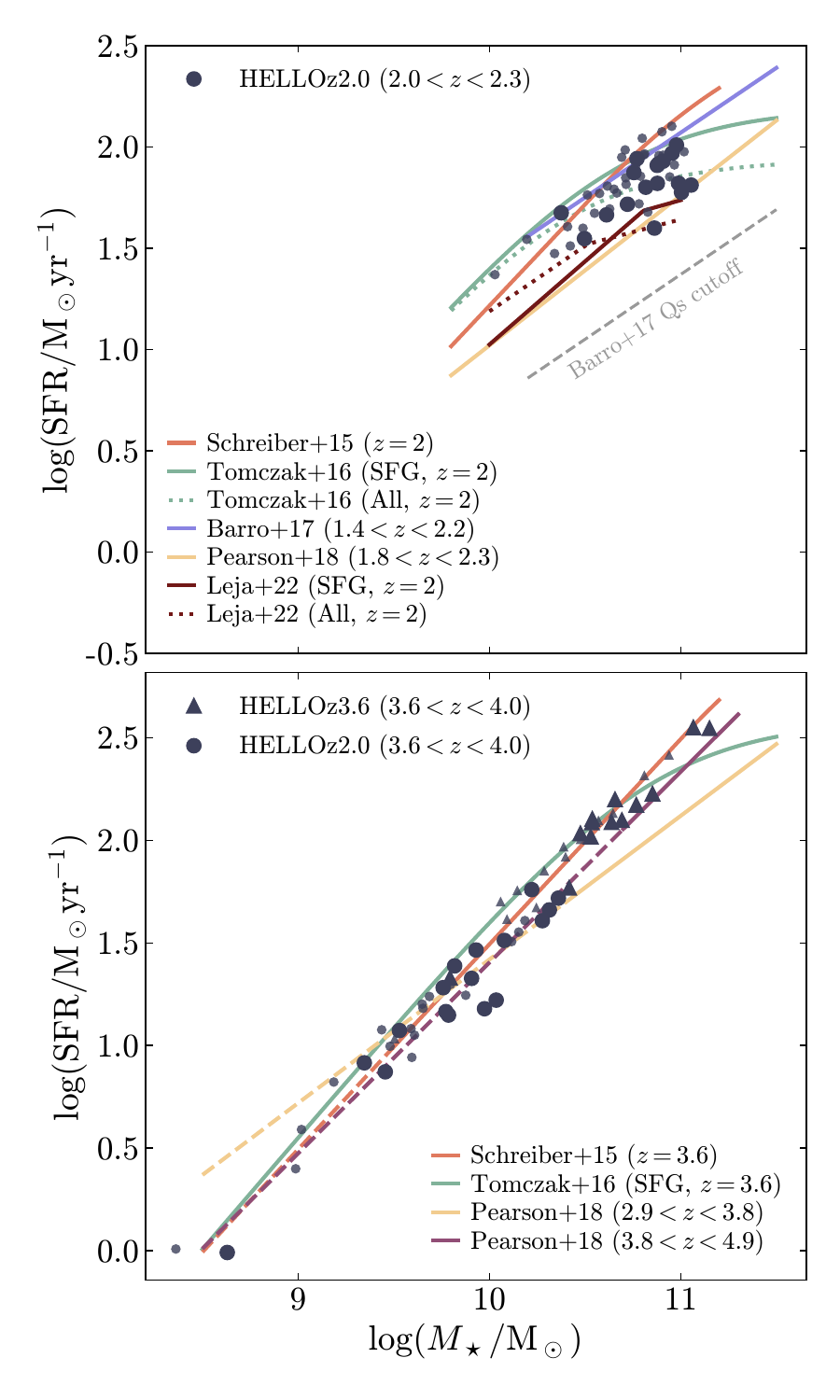}
\caption{
Stellar mass versus SFR for HELLOz2.0 between redshifts of 2.0 and 2.3 (top; circles) and HELLOz3.6 between redshifts of 3.6 and 4.0 (bottom; triangles) compared to various relations found in the literature, indicated in the legend.
In the bottom panel, we add HELLOz2.0 progenitors in the same redshift range (circles). Large markers show the galaxies at $z=2.0$ and $z=3.6$, smaller ones show the respective progenitors at earlier times.
For the observed relations, a dashed line indicates an extrapolation to lower masses.
}
\label{fig:sfr_mstar}
\end{figure}

We compare the SFR--$M_\star$ relation of our galaxies with various observations in Fig.~\ref{fig:sfr_mstar}.
Unlike the last figure, only the three (two) final snapshots of HELLOz2.0 (HELLOz3.6) are shown in the top and bottom panels, respectively.
The concerned redshift ranges are $2.0 < z < 2.3$ and $3.6 < z < 4.0$, as indicated in the legends.
In addition, HELLOz2.0 snapshots at $3.6 < z < 4.0$ are added to the lower panel.
The shape of the points follows the same convention as before, i.e., galaxies from HELLOz2.0 are represented as circles and the ones from HELLOz3.6 as triangles, while the size indicates if the corresponding marker represents the galaxy (large) at the final redshift or a progenitor (small) at earlier times.

In the upper panel, our simulations are compared to the empirical SFR--$M_\star$ relations identified by \citet{Schreiber2015} (red), \citet{Tomczak2016} (green; continuous line for SFGs and dotted for all galaxies), \citet{Barro2017} (purple), \citet{Pearson2018} (yellow), and \citet{Leja2022} (dark red; continuous line for SFGs and dotted for all galaxies).
All these relations together cover redshifts spanning from $z=1.8$ to $z=2.3$.
We furthermore display the quiescent cutoff from Barro et al., defined as 0.7\,dex below their SFMS, shown here with a dashed grey line.
In the lower panel, the colours follow the same convention, with the addition of a second redshift bin from Pearson et al.\ (dark purple).
Together, these observed relations cover redshifts ranging from $z=2.9$ to $z=4.9$.

The upper panel of Fig.~\ref{fig:sfr_mstar} exhibits good agreement with observations.
Specifically, our simulations nicely embrace the curve from \citet{Tomczak2016} (all galaxies) and seem consistent with a turnover and flattening, which we attempt to quantify at the end of this section.
HELLOz2.0 slightly undershoots \citet{Schreiber2015} and \citet{Barro2017}, as well as the sample restricted to star-forming galaxies (SFGs) from \citet{Tomczak2016}. Most of the SFMS arising from theory tend to lie systematically lower than the observed main sequences around $z\sim$ 1--2 \citep[][]{Dutton2010, Sparre2015, Furlong2015, Sommerville2015_mnras, Dave2019, Donnari2019}.
In the NIHAO introductory paper, however, \citet{Wang2015} showed that their galaxies do not suffer from this discrepancy. Since HELLO shares the same code, it is thus likely that we are observing our galaxies at the knee of the SFMS, indicating that some of them might have begun their quenching phase and are transitioning away from it, as we will show at the end of this section.
Nevertheless, all our galaxies are well above the quiescent cutoff from \citet{Barro2017} and are therefore all considered to be still star-forming, having SFRs well in the order $10\,\Msun\,\mathrm{yr}^{-1}$.

Finally, our galaxies seem generally consistent with the relations from \citet{Pearson2018} and \citet{Leja2022}, albeit overpredicting the normalisation. The former argue that their results lie below other SFMS because of the method used to determine SFRs. These are obtained from full SED fitting, which attributes some of the IR luminosity to the older stellar population, thereby reducing the fraction imputed to young stars.
\citet{Leja2022} make similar arguments for their SFMS and deduce that the discrepancy between observations and theory originates from an overestimation of SFRs by IR and UV+IR methods.
We refer the interested reader to \citet{Popesso2023}, who compile over one hundred SFMS from the literature, for an extensive review and further discussions.

Moving on to the lower panel of Fig.~\ref{fig:sfr_mstar}, HELLO galaxies in the redshift range $3.6 < z < 4.0$ reproduce the observed SFMS extremely well across the two orders of magnitude of masses probed.
The only exception is \citet{Pearson2018} for redshifts $2.9 < z < 3.8$, whose slope is shallower than the two other predictions, as well as their higher redshift bin sample.
At these early cosmic times, the turnover is expected to be virtually non-existent, although \citet{Tomczak2016}\footnote{
Here, we only show the relation from Tomczak et al.\ for SFGs, since the difference with the `all' sample is relatively small and would needlessly clutter the figure.
} found a significant flattening still occurring at about $M_\star \sim 10^{11}$\,\Msun\ \citep[see also,][]{Popesso2023}.
While we lack a significant number of galaxies with stellar masses $\gtrsim 10^{11}\,\Msun$, our galaxies visually do not appear to exhibit any flattening.

Note that the lack of any quenched galaxy in our sample restricts our results to star-forming galaxies.
This nevertheless does not invalidate these results, as all observed SFMSs shown in Fig.~\ref{fig:sfr_mstar} exclude quiescent galaxies from their samples, unless stated otherwise.

%%%%%%%%%%%%%%%%%%%%%%%%%%%%%%%%%%%%%%%%%%%%%%%%%%%
\subsubsection{Quantifying variation in slope and consistency with observed SFMS}\label{subsubsec:sfms_slope}
%%%%%%%%%%%%%%%%%%%%%%%%%%%%%%%%%%%%%%%%%%%%%%%%%%%
\begin{figure*}
\centering
\includegraphics[clip=true, trim= 4mm 4mm 4mm 4mm, width=\linewidth]{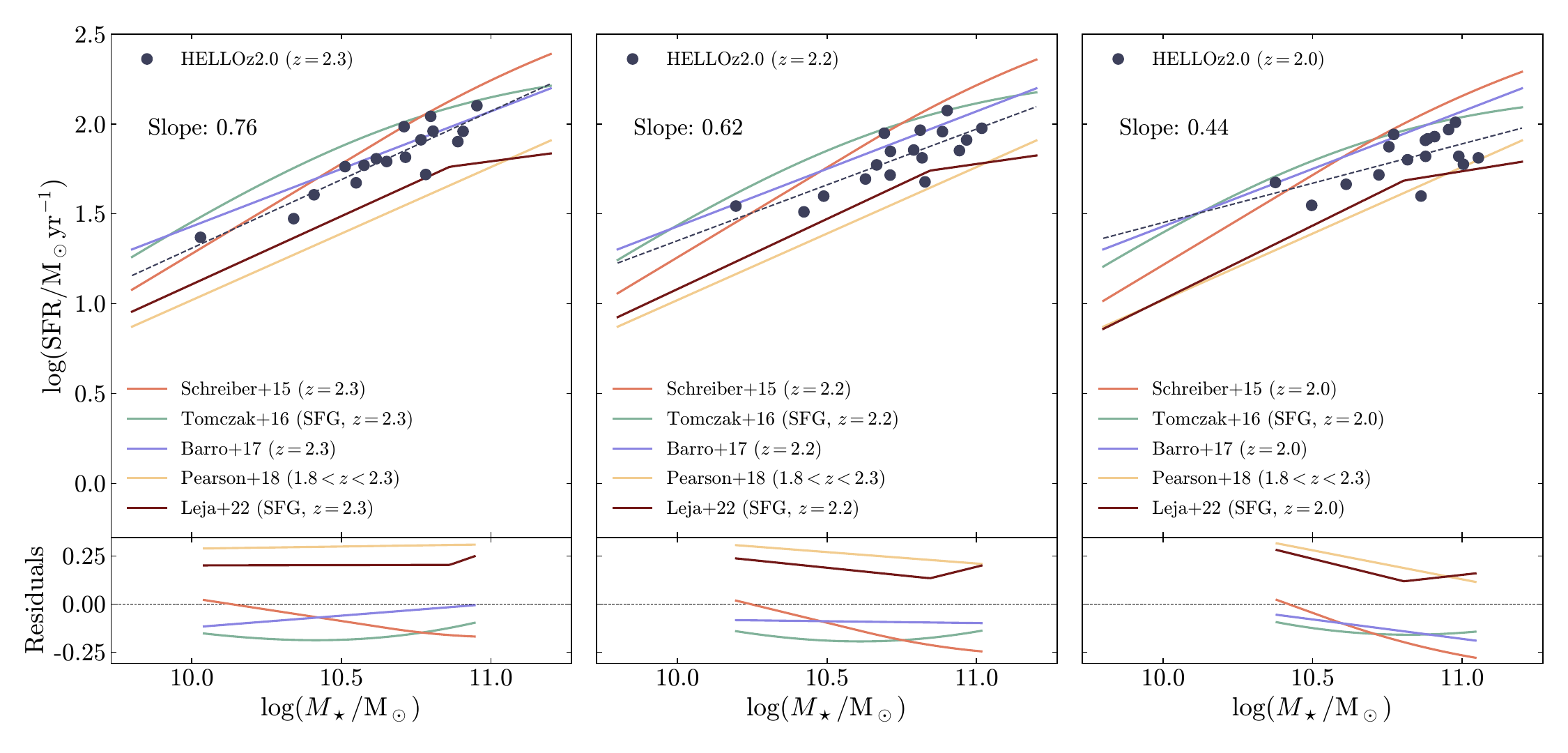}
\caption{SFMS of HELLOz2.0 galaxies at $z=2.3$ (left), $z=2.2$ (middle) and $z=2.0$ (right) compared to the same observations as Fig.~\ref{fig:sfr_mstar}. Each point represents the snapshot of a single galaxy at the respective redshift. We fit a single power law to our data, shown as a dashed line. The slope from the fit is indicated in the upper left of each panel. The residuals $\left\langle \log\,\mathrm{SFR}_{\mathrm{sim}} \right\rangle - \left\langle \log\,\mathrm{SFR}_{\mathrm{obs}} \right\rangle$ in the mass range covered by our simulations are shown below each panel.}
\label{fig:shmr_slope_z2}
\end{figure*}

\begin{figure*}
\centering
\includegraphics[clip=true, trim= 4mm 4mm 4mm 4mm, width=\linewidth]{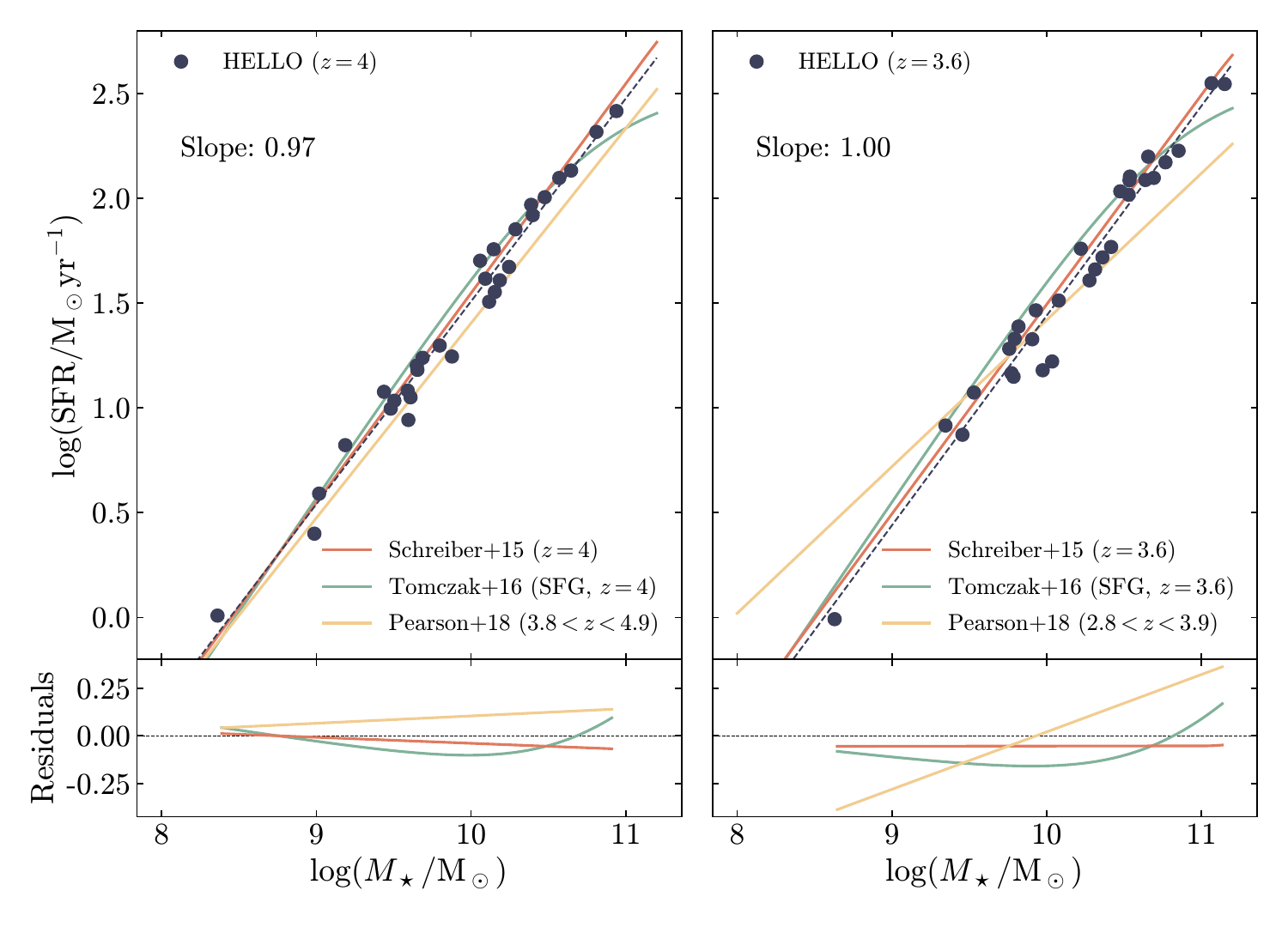}
\caption{SFMS of HELLO (HELLOz3.6 + HELLOz2.0) galaxies at $z=4.0$ (left), and $z=3.6$ (right) compared to the same observations as Fig.~\ref{fig:sfr_mstar}. Each point represents the snapshot of a single galaxy at the respective redshift. We fit a single power law to our data, shown as a dashed line. The slope from the fit is indicated in the upper left of each panel. The residuals $\left\langle \log\,\mathrm{SFR}_{\mathrm{sim}} \right\rangle - \left\langle \log\,\mathrm{SFR}_{\mathrm{obs}} \right\rangle$ in the mass range covered by our simulations are shown below each panel.}
\label{fig:shmr_slope_z3}
\end{figure*}

Despite a limited number of galaxies and a narrow range of masses covered, we attempt to quantify i) if our data are consistent with a flattening of the main sequence towards high \Mstar, and ii) how our fit compares to the observed relations in the (limited) range covered by our simulations.

To achieve this, we employ the method of ordinary least squares to fit a simple power law $\mathrm{SFR} \propto M_\star^{\alpha}$, where $\alpha$ represents the slope of the fit, and we compute the residuals between our SFMS and the observed ones. The residuals are calculated over the mass range covered by our galaxies and we extrapolate the observed relations if required.  We perform this exercise for a total of five different snapshots. For HELLOz2.0, we fit the SFMS at $z=2.3$, $z=2.2$, and $z=2.0$, using the respective snapshots of each galaxy.
For HELLOz3.6, we fit the SFMS at $z=4.0$ and $z=3.6$ by further adding the ancestors of HELLOz2.0 at these redshifts.

The results are shown in Figs.~\ref{fig:shmr_slope_z2} and \ref{fig:shmr_slope_z3}.
The figures are similar to Fig.~\ref{fig:sfr_mstar}, but now each panel only includes galaxies at fixed redshift.
Each point represents a single galaxy and we do not visually distinguish between HELLOz2.0 and HELLOz3.6, as in Fig.~\ref{fig:sfr_mstar}.
The dashed line represents the fit performed on those points and the residuals w.r.t.\ observed relations are shown in the lower panels.

Around $z\sim2$, our galaxies exhibit a SFMS with a relatively steady decline in its slope by 20--30~per~cent in $\sim\!200$\,Myr intervals, representing a total decline of roughly 40~per~cent between $z=2.3$ and $z=2.0$, or a time span of $\sim\!400$\,Myr.
These results indicate that around $z \sim 2$, the SFR of HELLO galaxies start to plateau, leading to a turnover in the slope of the main sequence. This trend is also visible in the residuals between our fit and the different relations from the literature. Specifically, the residuals exhibit a slope matching the one from \citet{Leja2022} before the turnover at $z=2.3$ and adjusting towards the high-mass slope at $z=2$.
Comparing now to \citet{Tomczak2016}, who incorporate the flattening in a smooth curve (as opposed to a broken power law), all three redshifts probed display residuals with a relatively constant shape, indicating that our SFMS evolves similarly, albeit lying $\sim\!0.1$\,dex below.
Aggregating all observations, our fit remains within $\sim\!0.25$\,dex of the observed ones.

From $z=4.0$ to $z=3.6$ on the other hand, the locus of our SFMS retains a slope consistent with 1, over the $\sim\!200$\,Myr separating both redshifts. Moreover, the residuals show excellent agreement with \citet{Schreiber2015}, \citet{Tomczak2016}, and \citet{Pearson2018} at $z=4$.
Transitioning to $z=3.6$, the agreement of our fit remains remarkably consistent with \citet{Schreiber2015} and \citet{Tomczak2016}, but shows discrepancy with \citet{Pearson2018} at the final redshift, as discussed previously.

We stress, however, that the turnover at $z\sim2$ appears in the \textit{time} evolution of the SFMS stemming from the same galaxies at three different redshifts.
It is not the exact equivalent of the flattening observed, for example, by \citet{Tomczak2016} or \citet{Leja2022}, as these relations are obtained from one sample of independent galaxies over a certain mass and redshift range.
Our results indicate, however, that while at $z\sim2$ our simulations do show a significant decline in the slope of the SFMS, this trend is not present at $z\sim3.6$ and the slope remains constant around unity.
Our data still remains consistent with \citet{Tomczak2016}, but the lack of galaxies with $\Mstar \gtrsim 10^{11}\,\Msun$ prevents us from making any stronger claims.

%%%%%%%%%%%%%%%%%%%%%%%%%%%%%%%%%%%%%%%%%%%%%%%%%%%
\subsection{Size versus mass relation}\label{subsec:size_mass_relation}
%%%%%%%%%%%%%%%%%%%%%%%%%%%%%%%%%%%%%%%%%%%%%%%%%%%

\begin{figure*}
\centering
\includegraphics[clip=true, trim= 4mm 4mm 4mm 4mm, width=\linewidth]{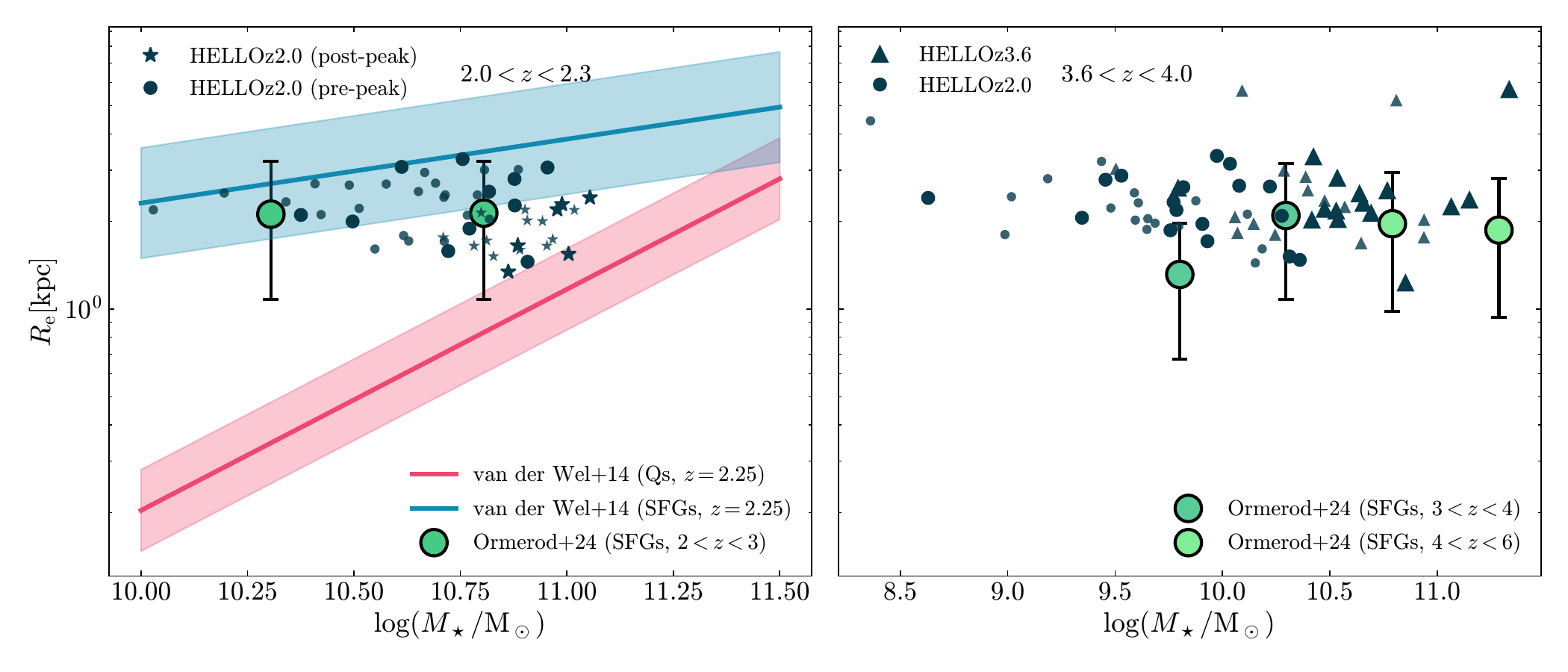}
\caption{
Size (effective radius, \Reff) versus (stellar) mass relation of HELLOz2.0 simulations between redshifts of 2.0 and 2.3 (left), and HELLOz2.0 and HELLOz3.6 between redshifts of 4.0 and 3.6 (right) compared to the observed relations determined by \citet{vanderWel2014} for quiescent (red) and star-forming (blue) galaxies, as well as the relation for star-forming galaxies obtained by \citet{Ormerod2024} using \textit{JWST} data (green circles with error bars).
The respective shaded regions and error bars depict the $\pm1\,\sigma$ scatter bounds.
On the left, HELLOz2.0 post-peak galaxies (large markers) and their progenitors (small markers) are marked as stars.
Similarly, pre-peak galaxies are depicted as circles.
On the right, the marker sizes follow the same convention and HELLOz2.0 are depicted as circles and HELLOz3.6 as triangles.
}
\label{fig:size_mass_relation}
\end{figure*}

Fig.~\ref{fig:size_mass_relation} shows the size-mass relation of HELLOz2.0 ($2.0 < z < 2.3$) galaxies plotted against the observed relation at $z=2.25$ inferred by \citet{vanderWel2014}, as well as the relation from \citet{Ormerod2024} (green circles with error bars) at $2<z<3$ on the left panel. We split our simulated galaxies into pre-peak (black dots) and post-peak (black stars) samples. The right panel shows HELLOz3.6 (triangles) and HELLOz2.0 (circles) galaxies between $z=3.6$ and $z=4.0$, compared to \citet{Ormerod2024} at $3<z<4$ and $4<z<6$.
Large markers denote galaxies at $z=2.0$ and $z=3.6$, while smaller ones correspond to progenitors at higher redshift. To calculate $\Reff$, we use the \texttt{pynbody.analysis.luminosity.half\_light\_r()} method from \textsc{\small PYNBODY} \citep{pynbody}, applied to all star particles within 20~per~cent of the virial radius.
This method calculates the radius at which half of the total luminosity in the given band (here $V$) is reached.

Starting with the left panel, the two observed relations that we included offer an interesting comparison to our simulations as they stem from \textit{HST} and \textit{JWST}, respectively.
Indeed, \citet{vanderWel2014} studied the evolution of the size-mass distribution over the redshift range $0 < z < 3$, using data from the 3D-HST \citep[][]{Brammer2012} and CANDELS \citep[][]{Grogin2011} surveys. On the other hand, \citet{Ormerod2024} made use of the latest data from the CEERS survey \citep[][]{Finkelstein2017, Finkelstein2023, Bagley2023}, using \textit{JWST} NIRCam imaging \citep[][]{Rieke2023}.

While in broad agreement with the SFG relation from \citet{vanderWel2014}, HELLOz2.0 pre-peak galaxies tend to underpredict the relation and this trend is accentuated with increasing stellar mass.
Moreover, post-peak galaxies are systematically outside the $1\sigma$ region and lie in between the relation from star-forming and quiescent galaxies.
Interestingly, our data seem to reconcile better with sizes inferred from \textit{JWST}, especially at the high-mass end.
Around $z \sim 4$, on the right panel, HELLO simulations exhibit sizes consistent with \citet{Ormerod2024}, although towards lower stellar masses ($\Mstar \lesssim 10^{10}\,\Msun$), our data shows some tension with the observations from \textit{JWST}.
The authors limit their sample to galaxies with $\log(\Mstar / \Msun) > 9.5$ which, combined with our own limited number of simulations, renders making additional claims difficult.
We further note that their sample lacks any SFG above $2 \times 10^{10} \,\Msun$ at $3 < z < 4$ and is the reason why we included the high-mass ($\sim$$10^{11}\,\Msun$) bins from their relation at $4 < z < 6$ with the caveat that the number of galaxies there is extremely limited and of the order one.

Despite the limitations mentioned above, these results offer interesting insights and we propose several explanations for what is seen in Fig.~\ref{fig:size_mass_relation}.

First, \citet{Arora2023} compared a multitude of scaling relations between NIHAO simulations and local galaxies from the MaNGA survey \citep[][]{Bundy2015, Wake2017} and found the high-mass simulations above $\logmstar \sim 10$ tend to lie below the locus of the observed size-mass relation.
A possible cause could be a relative weakening of the stellar feedback with respect to the galaxy mass. This weakening, in turn prevents over-cooling and thus efficient baryonic accretion towards the center of a galaxy \citep[see, e.g.,][]{McCarthy2012}.
This effect can potentially be balanced by a take over from AGN feedback, but the latter might not be strong enough yet (see \S\ref{sec:gas_and_feedback}).
With NIHAO and HELLO sharing the same hydrodynamical code, we argue that similar effects could be at play in our case.

Second, we do not apply any magnitude cutoff in the calculation of the effective radius.
\citet{Ma2018} have shown that the observed size of a galaxy can significantly depend on the surface brightness limit of a given observation campaign.
Taken at face value, this limiting factor tends however to an underestimation of $R_{\mathrm e}$ rather than the opposite.\footnote{
We also do not apply a conversion of our calculated \Reff\ to their equivalent at a rest-frame wavelength of 5000\,\AA.
}

Third, our calculation of $\Reff$ is `dust-free'.
Recent works employing simulations of galaxies at the epoch of reionisation \citep[][]{Marshall2022, Roper2022} show that dust attenuation can lead to significantly larger observed $\Reff$. 
Moreover, as pointed out by \citet{Ormerod2024}, \textit{JWST} is highly effective at measuring galaxies in the rest-frame optical range, where dust effects are minimised.
This could explain why our simulations, which are devoid of the effects of dust, appear smaller than \textit{HST} observation and more on par with \textit{JWST} ones.

%%%%%%%%%%%%%%%%%%%%%%%%%%%%%%%%%%%%%%%%%%%%%%%%%%%
\subsection{Surface density scaling relations}\label{subsec:surf_den}
\begin{figure}
\centering
\includegraphics[clip=true, trim= 3mm 6mm 6mm 6mm, width=\columnwidth]{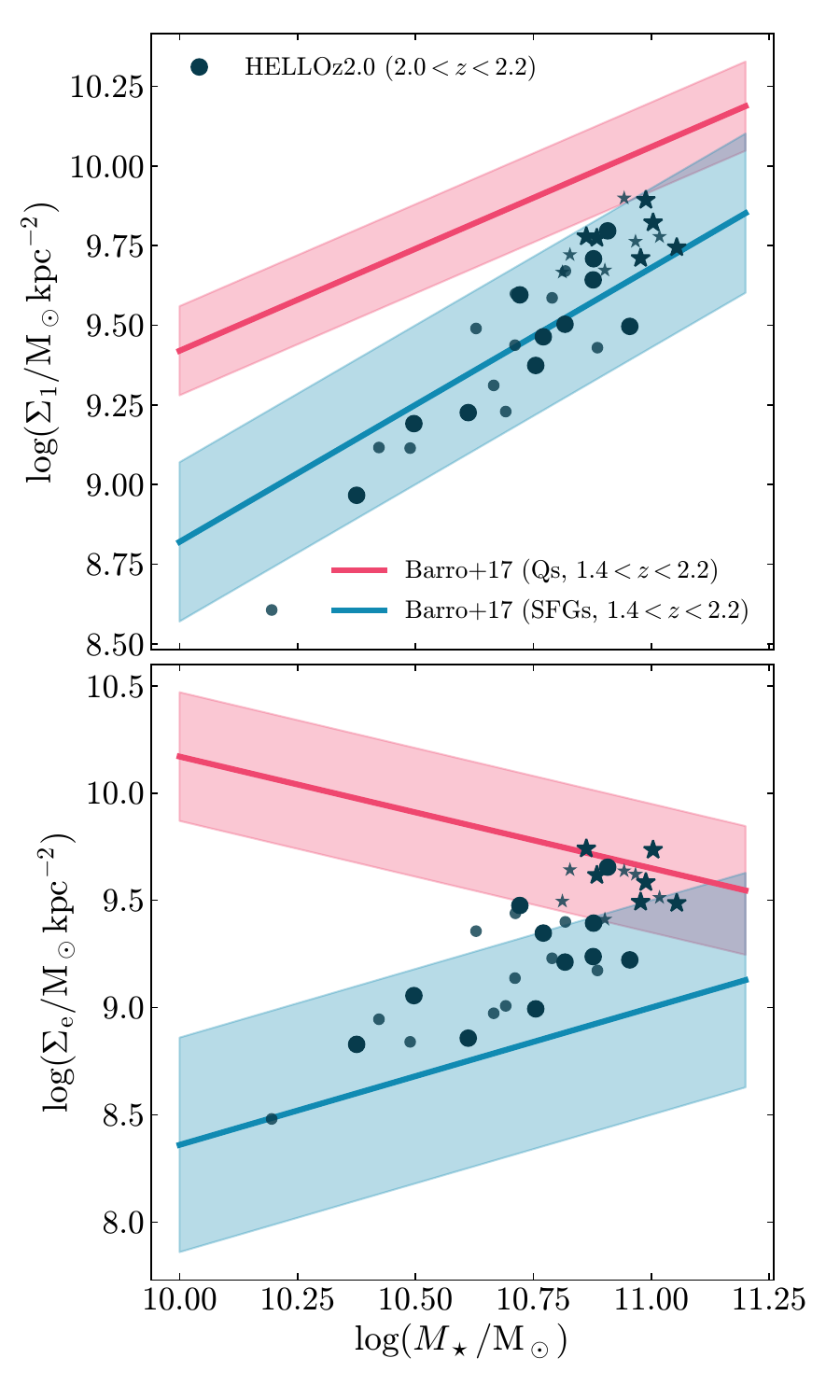}
\caption{
Surface density versus stellar mass relations of HELLOz2.0 simulations between redshifts of 2.0 and 2.2, compared to the observed relations determined by \citet{Barro2017} for quiescent (red) and star-forming (blue) galaxies.
The upper panel shows the relation for the central (within 1\,kpc) density ($\Sigma_1$) and the lower one for the effective (within $\Reff$) surface density ($\Sigmae$).
The shaded regions depict the $\pm1\,\sigma$ scatter.
HELLOz2.0 post-peak galaxies (large) and their progenitors (small) are marked as stars.
Similarly, pre-peak galaxies are depicted as circles.
}
\label{fig:surf_den_relation}
\end{figure}

%%%%%%%%%%%%%%%%%%%%%%%%%%%%%%%%%%%%%%%%%%%%%%%%%%%
We now turn our attention to the surface density scaling relations, namely the correlations between the surface mass density $\Sigma_r = M_\star(<r) / 4\pi r^2$, and the stellar mass for a given radius $r$.
In what follows, we calculate $\Sigma_{1}$ and $\Sigma_\mathrm{e}$, representing the densities enclosed within radii of 1\,kpc and the effective radius $\Reff$, respectively. 
In Fig.~\ref{fig:surf_den_relation}, we investigate how our $z=2$ sample compares to \citet{Barro2017} in the surface density to stellar mass relation.
The upper and lower panels show $\Sigma_1$ and $\Sigma_\mathrm{e}$ against $M_\star$, respectively.
We use snapshots covering the $2 < z <2.2$ range and mark galaxies belonging to the post-peak subsample as stars.
The red and blue lines represent a power law fit to a selection of massive galaxies from the CANDELS GOODS-S catalogue \citep[][]{Guo2013}{}{} for quiescent (red) and star-forming (blue) galaxies, and the shaded regions are the $\pm1\,\sigma$ scatter around the relations.
In their paper, Barro et al.\ found that both relations have been in place since $z \sim 3$ with practically no change in slope or scatter ever since.

HELLO galaxies agree with the central density relation for SFGs remarkably well within 1\,kpc, both following its slope and scatter.
Nevertheless, they overpredict \Sigmae, a feature accentuated at higher masses and especially for the post-peak galaxies, which are well within the Barro et al.\ quiescent relation.
This is not surprising when comparing to Fig.~\ref{fig:size_mass_relation}.
Indeed, both observed samples stem from the same survey and instruments, and since \Sigmae\ directly depends on \Reff, smaller galaxies, on average, naturally exhibit higher effective surface densities.
We already discussed in \S\ref{subsec:size_mass_relation} some likely explanations for the differences seen between observations and simulations, and we thus conclude that the same justifications apply here.

%%%%%%%%%%%%%%%%%%%%%%%%%%%%%%%%%%%%%%%%%%%%%%%%%%%
\section{Gas availability and AGN feedback}\label{sec:gas_and_feedback}
%%%%%%%%%%%%%%%%%%%%%%%%%%%%%%%%%%%%%%%%%%%%%%%%%%%
In this section, we want to expand on the two main observations regarding SFR that can be drawn from HELLO simulations, namely that i) similar mass galaxies at higher redshift show higher SFR and ii) some galaxies at $z=2$ seem to have started their transition towards quiescence, while the others still exhibit raising SFHs.
Specifically, we investigate the possible sources and mechanisms leading to these differences, focusing on cold gas availability and AGN feedback.

%%%%%%%%%%%%%%%%%%%%%%%%%%%%%%%%%%%%%%%%%%%%%%%%%%%
\subsection{Gas temperature and density}\label{subsec:gas_temp_dens}
%%%%%%%%%%%%%%%%%%%%%%%%%%%%%%%%%%%%%%%%%%%%%%%%%%%

\begin{figure}
\centering
\includegraphics[clip=true, trim= 2mm 3mm 2mm 2mm, width=\columnwidth]{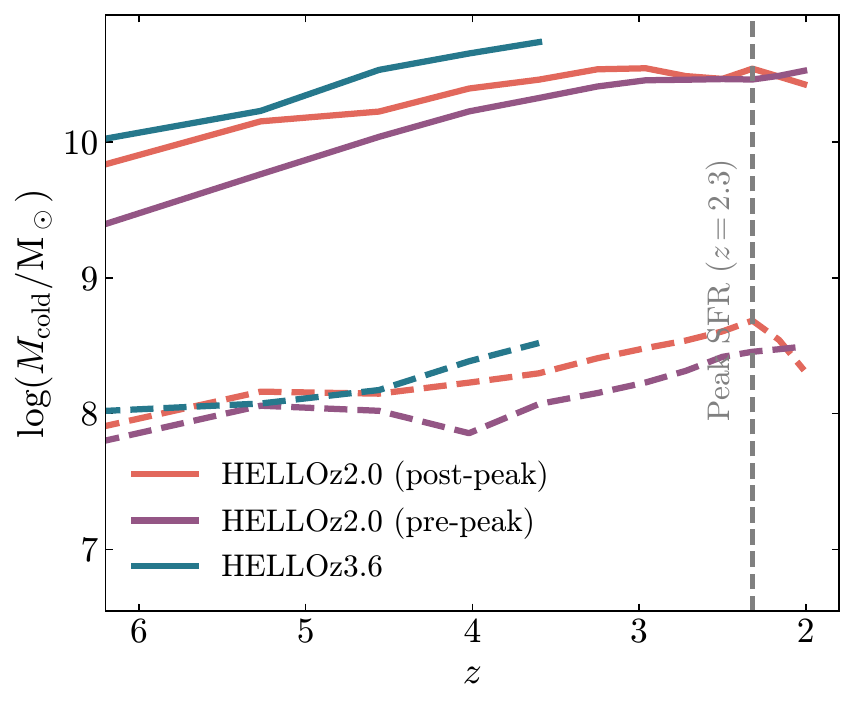}
\caption{
Median cold ($T < 15000$\,K) gas mass time evolution for HELLOz2.0 post-peak (orange), HELLOz2.0 pre-peak (purple), and HELLOz3.6 (turquoise) galaxies since $z=6$.
The continuous curves represent $M_{\mathrm{cold}}$ within 0.2\,\Rtwohun, while the dashed ones are for the central cold gas mass within 1\,kpc.
The grey dashed line indicates the epoch when the post-peak sample reaches its median peak SFR.
}
\label{fig:gas_mass_evolution}
\end{figure}

We begin by plotting the median cold ($T < 15000$\,K) gas mass time evolution of our galaxies in Fig.~\ref{fig:gas_mass_evolution}.
The continuous lines show the mass of cold gas within $0.2\,\Rtwohun$ for HELLOz2.0 post-peak (orange), HELLOz2.0 pre-peak (purple), and HELLOz3.6 (turquoise) galaxies.
The dashed lines represent the cold gas mass within 1\,kpc and the vertical dashed grey line indicates where the post-peak sample reaches the peak median SFR at $z=2.3$.
All samples go through a similar smooth evolution overall, with the amount of (central) cold gas mass increasing steadily over time and starting to saturate below $z \sim 3$ in HELLOz2.0.

At $z=2$, both the pre-peak and post-peak galaxies have roughly the same amount of cold gas available, but with an opposite trend.
While (central) cold gas mass is still increasing in the pre-peak sample, it is decreasing in the pre-peak sample after $z=2.3$, which corresponds to when post-peak galaxies reach their median peak SFR.
At their final redshift, HELLOz3.6 galaxies possess roughly twice as much median cold gas within 20~per~cent of $\Rtwohun$, but a similar central amount compared to HELLOz2.0.
A list of the exact values for each galaxy at the final redshift can be found in Appendix~\ref{app:hello_quantities}.

\begin{figure*}
\centering
\includegraphics[clip=true, trim= 6mm 6mm 6mm 6mm, width=\textwidth]{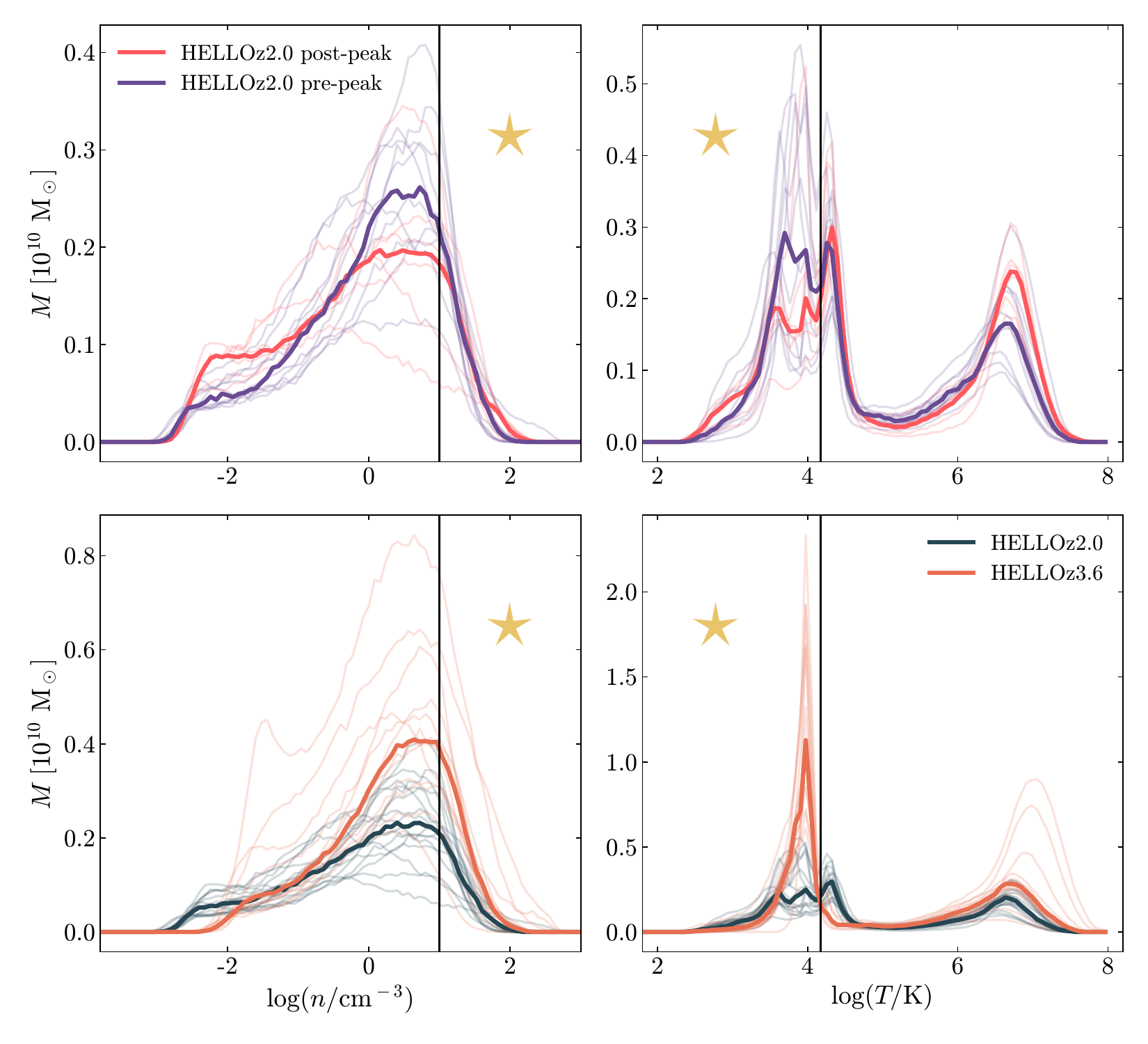}
\caption{
Gas distributions of our simulations, binned in log-density (left column) and log-temperature (right column) for particles within $0.2\,\Rtwohun$.
The upper row compares HELLOz2.0 post-peak (red) with pre-peak (purple) galaxies, and the lower row compares the whole samples of HELLOz2.0 (dark green) with HELLOz3.6 (brown).
Translucent curves represent individual galaxies and their median is plotted as a solid opaque curve.
In each panel, the vertical black line indicates the location of the density and temperature threshold, and the yellow star indicates the values allowing star formation.
}
\label{fig:gas_temp_dens}
\end{figure*}

Gas temperature is only half of the story in the ability of a galaxy to form stars; the gas also needs to reach a density of $n \geq 10$\,cm$^{-3}$ in our case.
Fig.~\ref{fig:gas_temp_dens} shows the distribution of gas mass within density (left panels) and temperature (right panels) bins, calculated within $0.2\,\Rtwohun$.
The upper row compares HELLOz2.0 pre-~and post-peak, while the lower one compares between the whole of HELLOz2.0 and HELLOz3.0 samples.
In each panel, the thick lines delineate the median of the lighter semi-transparent lines, representing individual galaxies, the vertical black line denotes where the threshold in $n$ and $T$ for star formation is, and the yellow star indicates the region (above or below the respective threshold) where star formation is allowed to take place.
Each curve is computed from their respective final snapshots.

Starting with temperature on the right, the two panels confirm what we observed in Fig.~\ref{fig:gas_mass_evolution}, namely that post-peak galaxies show slightly less, but a similar amount of available cold gas to form stars at the final redshift.
On the other hand, simulations at $z=3.6$ have clearly more cold gas at their disposal.
Individually, there is an apparent distinction between HELLOz2.0 and HELLOz3.6, meaning that not only the median values are different, but also individual galaxies at $z=3.6$ have systematically more cold gas than their $z=2.0$ counterpart.
This is unlike pre-~and post-peak, which display more overlapping curves.

Moving to the distribution in density (left panels), the findings are analogous between both redshifts.
HELLOz3.6 reveals a higher amount of dense gas, both individually and in the median.
At $z=2.0$, however, the pre-~and post-peak galaxies exhibit an overlapping median distribution of dense gas.
Interestingly though for densities $1 \lesssim n \lesssim 10$\,cm$^{-3}$, the median distributions show the largest disparities, indicating that pre-peak galaxies contain larger reservoirs of gas close to the star formation density threshold, thus allowing them to potentially sustain further SFR growth relative to post-peak galaxies.

We quantify this by calculating the mass of gas with both $n > 1$\,cm$^{-3}$ and $T < 10^5$\,K, as well as the corresponding fraction relative to the total mass of gas inside a galaxy.
We find for post-peak and pre-peak galaxies $\log(M_{\mathrm{cold, dense}}\,/\,\Msun) = 10.3 \pm 0.1$ and $10.4 \pm 0.1$, corresponding to $f_{\mathrm{cold, dense}} = 0.28_{-0.06}^{+0.04}$ and $0.35_{-0.04}^{+0.05}$, respectively.
Thus on average, pre-peak galaxies have $\sim\!25$~per~cent more cold and dense gas within $0.2\,\Rtwohun$, which also represents a larger fraction of the total gas.
Comparing the full HELLOz2.0 sample with HELLOz3.6, we obtain $\log(M_{\mathrm{cold, dense}}\,/\,\Msun) = 10.4 \pm 0.1$ and $\log(M_{\mathrm{cold, dense}}\,/\,\Msun) = 10.6 \pm 0.1$, translating to fractions of $f_{\mathrm{cold, dense}} = 0.34 \pm 0.06$ and $f_{\mathrm{cold, dense}} = 0.39_{-0.04}^{+0.01}$, respectively. 
HELLOz3.6 galaxies thus contain $\sim\!65$~per~cent more cold and dense gas, representing a larger fraction of total gas as well.

\begin{figure*}
\centering
\includegraphics[clip=true, trim= 2mm 3mm 2mm 2mm, width=\textwidth]{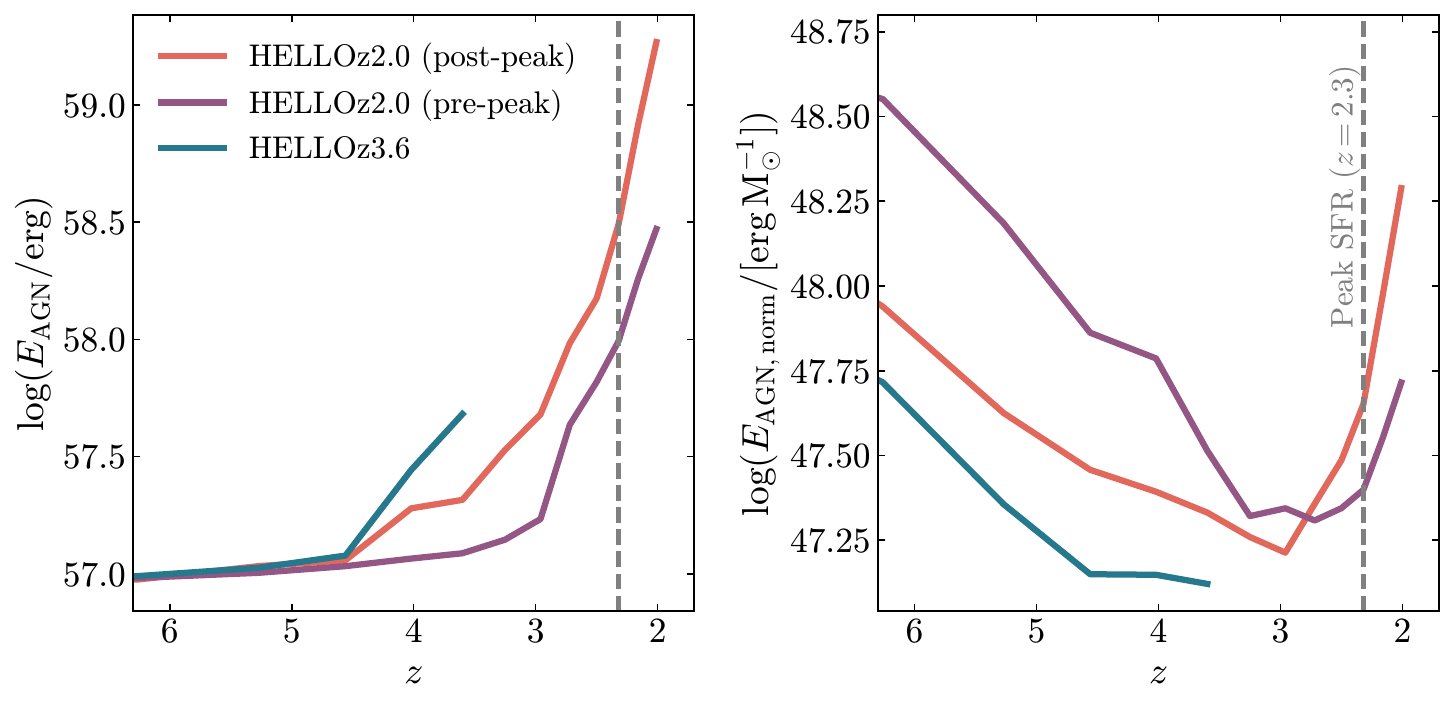}
\caption{
The left panel displays the median cumulative AGN feedback energy time evolution for HELLOz2.0 post-peak (orange), HELLOz2.0 pre-peak (purple), and HELLOz3.6 (turquoise) galaxies since $z=6$.
The grey dashed line indicates the epoch when the post-peak sample reaches its median peak SFR.
The right panel shows the same quantity normalised by the stellar mass.
The normalisation is applied to each galaxy prior to computing the median.
}
\label{fig:agn_cumulative}
\end{figure*}

%%%%%%%%%%%%%%%%%%%%%%%%%%%%%%%%%%%%%%%%%%%%%%%%%%%
\subsection{AGN feedback}\label{subsec:agn_feedback}
%%%%%%%%%%%%%%%%%%%%%%%%%%%%%%%%%%%%%%%%%%%%%%%%%%%

We now turn our attention towards the energy\footnote{According to Eq.~\ref{eq:agn_feedback}, the feedback energy is a direct proxy for BH mass, and thus accretion.} released by a central AGN and investigate if we can observe signs of its potential role in regulating its host galaxy's SFR.
Fig.~\ref{fig:agn_cumulative} contains the time evolution of the median cumulative AGN feedback energy released (left) and the same quantity normalised by the stellar mass at each redshift (right), i.e., $E_\mathrm{AGN,norm}(z) = E_\mathrm{AGN}(z) / M_{\star}(z)$.
The colour code in both panels follows the same convention as Fig.~\ref{fig:gas_mass_evolution}.
The three curves in the left panel exhibit a similar evolution, with a sudden exponential growth beginning between $z \sim 4.5$ and $z \sim 3$.
At their respective final redshifts, HELLOz3.6 galaxies have released a total of $\log(E_\mathrm{AGN}/\mathrm{erg})=57.7$ of AGN feedback energy, which is less than their counterparts at z=2.0, who reach $\log(E_\mathrm{AGN}/\mathrm{erg})=58.5$ and $\log(E_\mathrm{AGN}/\mathrm{erg})=59.3$ for pre- and post-peak samples, respectively.

When normalising by stellar mass (right panel), the following picture emerges.
During the initial phase of their evolution, the stellar mass growth of galaxies dominates over BH growth until the trend is reverted at a turnaround redshift of $z \sim 4$ for HELLOz3.6 and $z \sim 3$ for HELLOz2.0 galaxies.
In the second phase, BH growth dominates relative to stellar mass.
Notably unlike the samples at $z=2$, HELLOz3.6 \textit{does not} exhibit a dominating BH growth in its median curve yet.
Nevertheless, we looked at the most massive simulations individually, and their curve has indeed reverted.
We are therefore confident in our prediction that overall, the HELLOz3.6 galaxies would display a similar U-shaped curve if allowed to evolve further.

The right panel of Fig.~\ref{fig:agn_cumulative} suggests two main results.
First, assuming AGN feedback is (at least in some part) responsible for the quenching of galaxies, the value of $E_\mathrm{AGN, norm}$ alone is not sufficient to predict if a galaxy's SFR is about to weaken.
Indeed, the median value of $E_\mathrm{AGN, norm}$ for HELLOz2.0 at $z=6$ is similar to the value at $z=2.3$, which corresponds to peak SFR, having thus \textit{increased} during this time interval.
Second, given that both curves of pre-~and post-peak galaxies have already reverted, despite only the post-peak sample showing a decreasing SFR in the last few hundreds of Myr, we believe that the turnover in the $E_\mathrm{AGN, norm}$ evolution acts as a precursor to a slowing SFR and eventually quenching.
In other words, we predict that on average, the pre-peak sample is about to meet a similar fate as the post-peak one.

\begin{figure}
\centering
\includegraphics[clip=true, trim= 2mm 3mm 2mm 2mm, width=\columnwidth]{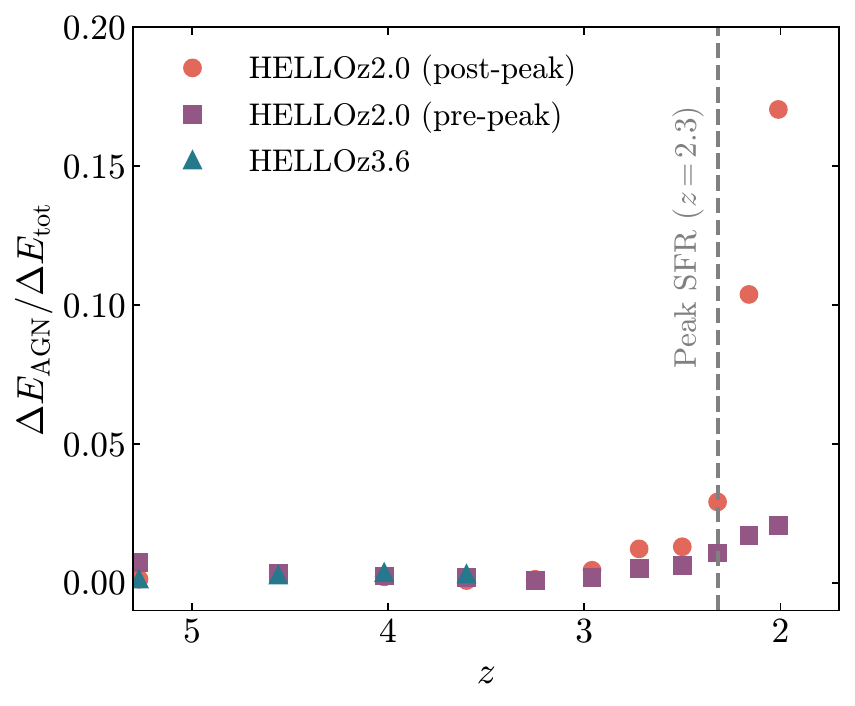}
\caption{
Fraction of energy released by the AGN relative to the total energy made up of AGNs, ESF, and SNe, in each snapshot.
HELLOz2.0 post-peak galaxies are represented as orange circles, HELLOz2.0 pre-peak as purple squares, and HELLOz3.6 as turquoise triangles, respectively.
The dashed grey line indicates where post-peak galaxies reach their median peak SFR at $z=2.3$. 
}
\label{fig:delta_agn_feedback}
\end{figure}

Fig.~\ref{fig:delta_agn_feedback} displays the relative amount of feedback energy released by the AGN with respect to the total feedback (AGN plus stars\footnote{See Appendix \ref{app:feedback_calculations} for details on how the stellar feedback is calculated.}) between each snapshot, i.e., in $\sim$200\,Myr timesteps. The HELLOz2.0 post-peak galaxies are represented as orange circles, HELLOz2.0 pre-peak as purple squares, and HELLOz3.6 as turquoise triangles, respectively. We purposefully avoid using a solid curve as before, to emphasise that Fig.~\ref{fig:delta_agn_feedback} \textit{does not} show the cumulative relative energy, but rather the energy emitted during each snapshot.

In our analysis, the interplay between stellar and AGN feedback arises as a pivotal factor in the energetic dynamics of galaxy evolution.
Up to $z=3$, stellar feedback is the predominant contributor to the energy output, completely overshadowing AGN feedback.
However, a notable shift occurs post $z=3$, where AGN feedback begins to assert a more significant role.
This trend is particularly pronounced in the HELLOz2.0 post-peak galaxies, where AGN feedback surges from 3 per cent to 17 per cent by $z=2$.
This uptick in AGN activity is linked to the concurrent decline in SFR, a dual phenomenon that underscores the complex interdependencies within galactic systems.

Although the increase in AGN feedback is less pronounced in pre-peak galaxies, it is noteworthy that its contribution steadily grows, even outpacing the rising SFR.
This observation is consistent with the trends noted in the right panel of Fig.~\ref{fig:agn_cumulative}.
Meanwhile, HELLOz3.6 galaxies are so efficient at forming stars that the fraction $\Delta E_{\mathrm{AGN}}$ relative to the total output $\Delta E_{\mathrm{tot}}$ remains negligible until the final redshift.
Yet, the leveling off observed in the right panel of Fig.~\ref{fig:agn_cumulative} suggests an impending shift in this trend, potentially aligning these galaxies with the evolutionary trajectory of HELLOz2.0 galaxies.
It is important to emphasise that while AGN feedback may be quantitatively lesser in overall energy contribution compared to stellar feedback, its localised nature, concentrated at the galactic centre, implies a disproportional impact on the immediate surroundings.

A potential limitation of our study lies in the dependence of our results on the Bondi model of BH accretion, where $\dot{M}_\bullet \propto M_\bullet^2$ (see Eq.~\ref{eq:bondi}).
As illustrated by \citet{Blank2019}, BH mergers are the primary catalysts for substantial increases in BH mass at early times, thus allowing for significant accretion and the onset of runaway growth.
This might be the reason behind the relatively late beginning of BH growth in HELLOz3.6 galaxies, which as a result, lack the time to accrete (and thus emit) as much as HELLOz2.0 BHs.
Furthermore, \citet{Soliman2023} demonstrated by using NIHAO simulations that the choice of the accretion model can significantly affect the evolution of a galaxy.
In the future, we are working on testing different implementations of AGN accretion and feedback (C.\ Cho et al, in prep.).

As was mentioned earlier in this work, our samples lack any quenched galaxy.
While we are dealing with low number statistics, this could nonetheless point to weaknesses in our AGN feedback modelling.
We refer the reader to \citet{Blank2022} for an in-depth discussion of the limitations of our model.
The authors found that NIHAO galaxies tend to cross the green valley in timescales shorter than observed.
This might allow our galaxies to quench \textit{later} but \textit{faster}, and could be a reason why we lack quenched galaxies in our samples.

%%%%%%%%%%%%%%%%%%%%%%%%%%%%%%%%%%%%%%%%%%%%%%%%%%%
\section{Summary and conclusion}\label{sec:discussion}
%%%%%%%%%%%%%%%%%%%%%%%%%%%%%%%%%%%%%%%%%%%%%%%%%%%
In this paper, we presented HELLO, a new set of high-resolution cosmological zoom-in simulations for massive MW-type galaxies at final redshifts of $z=2$ and $z=3.6$, respectively.
The hydrodynamics and sub-grid physics governing the evolution of these galaxies are identical to NIHAO \citep[][]{Wang2015, Blank2019} with the addition of LPF from \citet{Obreja2019} and a revised elemental feedback, as introduced by \citet{Buck2021}.
The HELLO sample used here consists of a total of 32 simulations with $\Mstar \sim 10^{10-11}\,\Msun$ and $N_\mathrm{part} \sim 10^6$ within their respective virial radii.
We summarise our results as follows:
\begin{enumerate}
    \item Our simulations reproduce the high SFRs measured in SFGs at 2--3\,Gyr after the Big Bang.
    Moreover, the time evolution of the slope in the SFMS of HELLOz2.0 shows a flattening between $z=2.3$ and $z=2.0$ and around $\logmstar \sim 10.5$, from 0.76 to 0.44.
    This flattening seems consistent with \citet{Tomczak2016} and \citet{Leja2022}, as demonstrated by the evolution of the residuals between our fit and observations.
    From $z=4.0$ to $z=3.6$, on the other hand, the slope remains consistent with one (Figs.~\ref{fig:shmr_slope_z2} and \ref{fig:shmr_slope_z3}).

    \item HELLO galaxies are in good agreement with the SFR--\Mstar\ relation found by various authors in the literature \citep[][]{Schreiber2015, Tomczak2016, Barro2017, Leja2022}, particularly at $z \sim 4$ where our simulations exhibit a remarkably tight distribution around the SFMS (Fig.~\ref{fig:sfr_mstar}).
    We recover the slopes of \citet{Pearson2018} and \citet{Leja2022} at $z=2.3$, and the slope of \citet{Barro2017} at $z=2.2$ (Fig.~\ref{fig:shmr_slope_z2}).

    \item Our study finds that HELLOz2.0 galaxies generally underpredict the size-mass relation observed for SFGs from \textit{HST} data \citep[][]{vanderWel2014}, especially at higher stellar masses where post-peak galaxies deviate significantly from the relation (Fig.~\ref{fig:size_mass_relation}, left).
    Possible reasons include relatively weaker stellar feedback with respect to \Mstar\ \citep[][]{McCarthy2012, Arora2023} and the onset of the quenching phase in post-peak galaxies. When compared to \textit{JWST}, however, our simulations are in excellent agreement with observations \citep[][]{Ormerod2024}, both around $z \sim 2$ and $z \sim 4$.
    In the latter and below $\log(\Mstar\,/\,\Msun) \sim 10.5$, though, our data exhibit some tension with \citet{Ormerod2024}, but the limited number of galaxies on our side and the mass floor on theirs makes any stronger conclusion challenging (Fig.~\ref{fig:size_mass_relation}, right panel).

    \item HELLO galaxies closely match the central density relation for SFGs within 1\,kpc \citep[][]{Barro2017}, in terms of both slope and scatter (Fig.~\ref{fig:surf_den_relation}).
    However, the simulations tend to overpredict the effective surface density, particularly for post-peak galaxies, aligning more with the quiescent relation.
    This overprediction follows naturally from our smaller sizes, as summarised in (iii).

    \item All samples exhibit a steady increase in cold gas mass on aggregate, with HELLOz2.0 beginning to saturate below $z \sim 3$ (Fig.~\ref{fig:gas_mass_evolution}).
    At $z=2$, pre-peak and post-peak galaxies have similar amounts of cold gas, but with contrasting trends: increasing in pre-peak and decreasing in post-peak galaxies after $z=2.3$, which coincides with the median peak SFR for post-peak simulations.
    HELLOz3.6 galaxies have about twice as much median cold gas mass overall as compared to HELLOz2.0 galaxies at their final redshift.

    \item Gas density analyses show that pre-peak galaxies have on average 25~per~cent more cold and dense gas than post-peak ones within $0.2\,\Rtwohun$. This value increases to more than 50~per~cent when comparing HELLOz3.6 and HELLOz2.0 as a whole.

    \item While stellar feedback dominates the feedback energy liberated, the central BHs of our galaxies have undergone substantial growth since $z \sim 4$ (Figs.~\ref{fig:agn_cumulative} and \ref{fig:delta_agn_feedback}).
    At their respective final redshifts, galaxies at $z=2$ have more massive BHs than their counterparts at $z=3.6$, with the post-peak sample having the most massive BHs.
\end{enumerate}

A general picture materialises from our results, which we describe as follows.
Beginning with galaxies at $z=2$, they undergo a relatively smooth progression along the SFMS, characterised by sustained SFR growth and cold gas accumulation. Post-peak galaxies, which are generally the most massive, consistently exhibit higher SFRs  until they reach their peak SFR around $z \sim 2.3$, coinciding with significant BH growth. As a result, this period witnesses the depletion of central gas, likely attributed to the central BH and stars consuming the gas, and in turn, emitting a substantial amount of energy, leading to a negative feedback loop a declining SFR.

Conversely, pre-peak galaxies have yet to achieve their peak SFRs.
They do not display a reduction in central cold gas reserves and their AGN activity is significantly less intense, as indicated by the lower feedback emission.
Nonetheless, we postulate that these galaxies are merely trailing behind their post-peak counterparts and will eventually mirror the latter in a few hundred Myr\footnote{
Analysis of subsequent snapshots shows that the SFRs of most of these galaxies start declining after $z=2$.
Past the target redshift, however, increasing risks of contamination from low-resolution DM particles in the high-resolution regions motivate our choice not to consider these results at face value in this work.
}, as indicated by their stellar masses and high SFRs \citep[see, e.g.][]{Peng2010}.

At $z=3.6$, galaxies possess larger reserves of cold and high-density gas, accounting for their higher SFRs.
They additionally host less massive BHs, whose feedback is substantially weaker than galaxies at $z=2$.
These differences could be attributed to the shorter cooling times prevalent at high redshift, allowing the gas to cool and reach the density threshold faster.
The gas, in turn, actively forms stars rather than feeding the central AGN.
We caution the reader that we could also be witnessing some limitations in our modelling of AGN accretion and feedback.
These will be further investigated in future works.

Our findings support the notion that the increase of AGN feedback is \textit{correlated} with declining SFRs, but we cannot claim that this correlation is a direct causal relation.
This does not preclude the simultaneous occurrence of intense AGN activity and high SFRs in SFGs \citep[][]{Hopkins2006}, as both post-~and pre-peak galaxies exhibit comparable median SFRs at their final redshifts, but follow inverse trajectories. The latter suggests that the exponential growth of central black holes and the ensuing feedback both precede the quenching process \citep[][]{DiMatteo2005, Springel2005b}.

It is worth mentioning that an abundance of theoretical works have investigated the large availability of cold gas and the peak in star formation at cosmic noon, and have attributed them to the prevalence of cold streams of gas able to deliver a constant and significant amount of fresh fuel directly to the central galaxy \citep[cold mode accretion;][]{Katz2003, Keres2005}.
These collimated streams are able to keep the gas unshocked as they penetrate the halo \citep[][]{Dekel2006, Dekel2009, Keres2009, Brooks2009, Voort2011, Aung2024}, even when the latter is above the threshold mass for a stable virial shock \citep[][]{Birnboim2003}. 
While at $z=3.6$ our galaxies exhibit a significant higher fraction of cold and dense gas compared to $z=2$, we reserve the exploration of its origin in a companion paper (Waterval et al., in prep.).

As it is, the HELLO simulations represent a meaningful step in progress toward an improved understanding of the nature of galaxies at high redshifts, and its forthcoming improvements will help to further advance its capabilities.

%\newpage 

\section*{Acknowledgements}
We are grateful to Alexander Knebe for his help regarding issues encountered with the \ahf, and to  Carlo Cannarozzo for devising the HELLO acronym. 
This material is based upon work supported by Tamkeen under the NYU Abu Dhabi Research Institute grant CASS.
TB’s contribution to this project was made possible by funding from the Carl Zeiss Foundation. A.O. has been funded by the Deutsche Forschungsgemeinschaft (DFG, German Research Foundation) – 443044596.
The authors  gratefully acknowledge the Gauss Centre for Supercomputing e.V.\ (\url{www.gauss-centre.eu}) for funding this project by providing computing time on the GCS Supercomputer SuperMUC at the Leibniz Supercomputing Centre (\url{www.lrz.de}) and the High Performance Computing resources at New York University Abu Dhabi.
%M.~P. acknowledges financial support from the European Union’s Horizon 2020 research and innovation program under the Marie Sk\l{}odowska-Curie grant agreement No.\ $896248$.

\section*{Data availability}
The data underlying this article will be shared on reasonable
request to the corresponding author.



\bibliographystyle{mnras}
\bibliography{ref} % if your bibtex file is called example.bib


%%%%%%%%%%%%%%%%%%%%%%%%%%%%%%%%%%%%%%%%%%%%%%%%%%

%%%%%%%%%%%%%%%%% APPENDICES %%%%%%%%%%%%%%%%%%%%%

%%%%%%%%%%%%%%%%%%%%%%%%%%%%%%%%%%%%%%%%%%%%%%%%%%
\appendix

\section{HELLO quantities at their final redshift}\label{app:hello_quantities}
\begin{landscape}
\begin{table}
    \centering
    \scalebox{0.9}{%
    \begin{tabular}{lrrrrrrrrrrrrrrrrrrrrr}
\toprule
     Name & $N_{\mathrm{tot}}$ & $N_{\mathrm{DM}}$ & $N_{\star}$ & $N_{\mathrm{gas}}$ & $R_{200}$ &  $R_{\mathrm{e}}$ & $M_{\mathrm{halo}}$ &  $M_{\mathrm{DM}}$ & $M_{\star}$ &  $M_{\mathrm{gas}}$ & $M_{\mathrm{cold}}$ & $M_{\mathrm{cold},1}$ &   $M_{\bullet}$ & $\mathrm{SFR}_{10}$ & $\mathrm{SFR}_{100}$ & $\Sigma_1$ & $\Sigma_{\mathrm{e}}$ &  $\Delta \mathrm{ESF}$ & $\Delta \mathrm{SNe\,II}$ \\
      & & & & & & & $\times 10^{12}$ & $\times 10^{12}$ & $\times 10^{10}$ & $\times 10^{10}$ & $\times 10^{12}$ & $\times 10^{8}$ & $\times 10^{6}$ & & & $\times 10^{9}$ & $\times 10^{9}$ & $\times 10^{59}$ & $\times 10^{57}$\\
      & & & & & [a] & [a] & [b] & [b] & [b] & [b] & [b] & [b] & [b] & [c] & [c] & [d] & [d] & [e] & [e]\\
\midrule
g3.08e12 & 1,943,368 &   732,154 &   752,300 &   458,914 & 142 & 1.5 & 2.84 & 2.48 & 10.09 &  7.31 &  2.60 &  2.07 & 34.42 &  55 &  60 & 6.64 & 5.43 &  3.73 &  2.87\\
g3.09e12 & 2,058,093 &   758,531 &   836,846 &   462,716 & 144 & 2.4 & 2.94 & 2.57 & 11.33 &  7.26 &  2.47 &  0.00 & 74.07 &  68 &  65 & 5.56 & 3.07 &  4.07 &  3.14\\
g3.00e12 & 1,501,948 &   626,696 &   433,000 &   442,252 & 135 & 1.9 & 2.43 & 2.12 &  5.91 &  8.33 &  3.45 &  4.11 &  0.51 &  99 &  87 & 2.91 & 2.23 &  4.13 &  3.18\\
g3.20e12 & 1,693,806 &   694,672 &   456,825 &   542,309 & 140 & 3.3 & 2.72 & 2.35 &  5.69 &  9.27 &  4.35 &  2.61 &  3.31 &  81 &  75 & 2.37 & 0.99 &  3.92 &  3.02\\
g2.75e12 & 1,778,286 &   624,856 &   720,877 &   432,553 & 135 & 2.3 & 2.45 & 2.12 &  9.74 &  8.77 &  4.34 &  2.08 & 24.14 &  66 &  66 & 7.82 & 3.84 &  3.96 &  3.05\\
g3.03e12 & 1,250,411 &   426,342 &   548,685 &   275,384 & 119 & 2.3 & 1.67 & 1.44 &  7.55 &  5.69 &  2.35 &  2.78 &  5.56 &  54 &  66 & 5.12 & 2.48 &  3.85 &  2.96\\
g3.01e12 & 1,729,071 &   683,692 &   563,963 &   481,416 & 139 & 1.7 & 2.66 & 2.32 &  7.67 &  7.41 &  2.72 &  4.36 &  3.02 &  84 &  83 & 5.94 & 4.14 &  4.80 &  3.69\\
g2.29e12 & 1,333,337 &   460,720 &   543,304 &   329,313 & 122 & 2.8 & 1.81 & 1.56 &  7.53 &  7.71 &  3.19 &  3.79 &  4.94 &  95 &  81 & 4.39 & 1.73 &  4.30 &  3.31\\
g3.35e12 &   964,316 &   441,712 &   175,744 &   346,860 & 120 & 2.1 & 1.72 & 1.50 &  2.37 &  6.69 &  3.25 &  2.71 &  1.05 &  53 &  47 & 0.93 & 0.67 &  2.47 &  1.90\\
g3.31e12 & 1,449,371 &   594,285 &   408,855 &   446,231 & 133 & 1.6 & 2.32 & 2.01 &  5.27 &  8.21 &  3.79 &  5.05 & 13.40 &  79 &  52 & 3.94 & 2.99 &  3.13 &  2.41\\
g3.25e12 & 1,643,064 &   674,577 &   412,909 &   555,578 & 139 & 3.1 & 2.65 & 2.29 &  4.09 &  8.11 &  4.55 &  2.26 &  1.19 &  58 &  46 & 1.68 & 0.72 &  2.67 &  2.06\\
g3.38e12 & 2,014,002 &   787,483 &   639,957 &   586,562 & 146 & 1.5 & 3.08 & 2.67 &  8.09 &  9.86 &  3.68 &  9.65 &  4.70 & 106 &  85 & 6.26 & 4.51 &  5.91 &  4.55\\
g3.36e12 & 1,209,250 &   529,555 &   251,774 &   427,921 & 128 & 2.0 & 2.08 & 1.79 &  3.14 &  4.43 &  1.95 &  2.16 &  2.22 &  31 &  35 & 1.55 & 1.14 &  1.85 &  1.42\\
g2.83e12 & 1,551,228 &   592,464 &   523,384 &   435,380 & 133 & 2.5 & 2.32 & 2.01 &  6.57 &  5.75 &  2.24 &  3.32 &  3.44 &  86 &  63 & 3.19 & 1.63 &  3.39 &  2.61\\
g2.63e12 & 1,682,819 &   596,827 &   661,422 &   424,570 & 133 & 3.1 & 2.35 & 2.02 &  9.01 &  7.85 &  3.34 &  3.12 &  1.36 &  93 &  93 & 3.14 & 1.67 &  5.14 &  3.96\\
g3.04e12 & 1,997,107 &   774,655 &   710,292 &   512,160 & 145 & 2.2 & 3.01 & 2.63 &  9.50 &  8.10 &  2.82 &  2.03 & 17.25 & 103 & 102 & 5.14 & 3.11 &  6.00 &  4.62\\
g2.91e12 & 1,445,948 &   541,666 &   542,915 &   361,367 & 129 & 1.3 & 2.11 & 1.84 &  7.29 &  4.33 &  1.36 &  4.22 & 13.63 &  40 &  40 & 6.01 & 5.51 &  2.38 &  1.83\\
\midrule
g2.47e12 & 1,259,900 &   567,753 &   247,017 &   445,130 &  87 & 2.2 & 2.21 & 1.92 &  3.39 &  8.16 &  3.94 &  2.61 &  0.44 & 121 & 104 & 1.22 & 0.92 &  5.09 &  3.91\\
g2.40e12 & 1,231,536 &   577,171 &   123,812 &   530,553 &  88 & 2.6 & 2.29 & 1.96 &  0.62 &  5.16 &  3.43 &  1.09 &  0.16 &  27 &  21 & 0.18 & 0.12 &  1.01 &  0.77\\
g2.69e12 & 1,493,797 &   650,082 &   332,760 &   510,955 &  91 & 2.3 & 2.54 & 2.20 &  4.53 & 10.95 &  5.47 &  4.14 &  0.28 & 178 & 158 & 1.89 & 1.17 &  7.13 &  5.49\\
g3.76e12 & 1,998,769 &   910,838 &   298,900 &   789,031 & 102 & 2.0 & 3.58 & 3.09 &  3.45 & 12.28 &  6.08 &  3.31 &  0.54 & 146 & 127 & 1.26 & 0.89 &  6.31 &  4.85\\
g4.58e12 & 1,616,009 &   662,886 &   417,102 &   536,021 &  92 & 2.6 & 2.60 & 2.25 &  5.86 & 11.22 &  5.44 &  4.89 &  1.12 & 165 & 149 & 3.65 & 1.60 &  7.66 &  5.89\\
g2.71e12 & 1,216,000 &   490,095 &   304,633 &   421,272 &  84 & 2.5 & 1.98 & 1.70 &  4.35 &  9.75 &  4.76 &  3.72 &  0.64 & 159 & 123 & 1.92 & 1.07 &  6.44 &  4.95\\
g2.49e12 & 1,427,551 &   647,600 &   227,010 &   552,941 &  91 & 2.2 & 2.55 & 2.19 &  2.99 &  9.72 &  4.83 &  2.76 &  0.14 & 113 & 108 & 1.30 & 0.87 &  5.13 &  3.95\\
g2.51e12 & 1,529,843 &   635,272 &   338,985 &   555,586 &  91 & 2.0 & 2.52 & 2.15 &  2.61 &  6.57 &  3.89 &  2.75 &  0.39 &  62 &  59 & 1.68 & 1.02 &  3.07 &  2.36\\
g7.37e12 & 3,526,490 & 1,465,334 &   866,819 & 1,194,337 & 120 & 2.2 & 5.76 & 4.97 & 11.64 & 20.65 &  8.87 & 11.06 &  1.42 & 407 & 355 & 6.05 & 3.14 & 16.88 & 12.99\\
g2.32e12 & 1,218,031 &   486,245 &   340,064 &   391,722 &  83 & 2.1 & 1.91 & 1.65 &  4.93 &  8.47 &  4.19 &  4.48 &  0.81 & 124 & 125 & 2.95 & 1.69 &  6.94 &  5.34\\
g2.96e12 & 1,605,548 &   716,314 &   258,236 &   630,998 &  95 & 2.8 & 2.84 & 2.45 &  3.43 & 12.71 &  6.63 &  2.10 &  0.26 & 159 & 122 & 0.99 & 0.62 &  5.83 &  4.48\\
g7.97e12 & 3,546,175 & 1,487,662 &   755,426 & 1,303,087 & 121 & 1.2 & 5.89 & 5.04 &  7.12 & 13.80 &  6.76 & 10.00 &  3.04 & 183 & 169 & 7.22 & 6.02 &  9.03 &  6.95\\
g9.20e12 & 5,318,120 & 2,230,919 & 1,317,484 & 1,769,717 & 138 & 2.4 & 8.74 & 7.56 & 14.15 & 25.68 & 10.61 &  1.21 & 66.69 & 351 & 352 & 9.39 & 4.07 & 19.82 & 15.25\\
\bottomrule
\end{tabular}
}
    \caption{
    Simulation quantities for HELLOz2.0 and HELLOz3.6 galaxies at their final redshifts of $z=2$ (top) and $z=3.6$ (bottom), respectively. Columns from left to right: galaxy name, total number of particles ($N_\mathrm{tot}$), number of DM particles ($N_\textrm{DM}$), number of star particles ($N_\star$), number of gas particles ($N_\mathrm{gas}$), virial radius ($\Rtwohun$), effective radius ($\Reff$), halo mass ($M_\mathrm{halo}$), DM mass ($M_\mathrm{DM}$), stellar mass ($M_\star$), gas mass ($M_\mathrm{gas}$), cold gas mass ($M_\mathrm{cold}$), cold gas mass within 1\,kpc ($M_\mathrm{cold,1}$), BH mass ($M_\bullet$), SFR averaged over the last 10\,Myr (SFR$_{10}$), SFR averaged over the last 100\,Myr (SFR$_{100}$), stellar surface density within 1 kpc ($\Sigma_1$), effective stellar surface density ($\Sigma_\mathrm{e}$), ESF energy released during the last 200\,Myr ($\Delta$ESF), and SNe\,II energy released during the last 200 Myr ($\Delta$SNe\,II).\\
    The different numbers of particles, as well as $M_\mathrm{halo}$ and $M_\mathrm{DM}$, are computed within $R_{200}$, while baryonic quantities are computed within $0.2\,\Rtwohun$, unless otherwise specified by the indices `1' or `e.'\\
    Units: [a] $\equiv$ [kpc], [b] $\equiv$ [$\Msun$], [c] $\equiv$ [$\Msun\,\mathrm{yr}^{-1}$], [d] $\equiv$ [$\Msun\,\mathrm{kpc}^{-2}$], and [e] $\equiv$ [erg].
    }
    \label{tab:my_label}
\end{table}
\end{landscape}

\section{Choice of galactic radius}\label{app:satellite_contamination}

In order to calculate various baryonic quantities such as $M_\star$, $M_\mathrm{gas}$, or SFR, we need to define the edges of the galaxy. 
Ideally, the chosen rule should i) be simple and consistent and ii) allow for as fair a comparison with observations as possible.
In this paper we chose a sphere enclosing 20~per~cent of \Rtwohun\ as our galactic radius for calculating galactic baryonic properties.
While somewhat arbitrary, this choice offers consistency across all galaxies within our sample, as well as with NIHAO and other simulations \citep[][]{Wang2015, Tacchella2016}.\footnote{
Some authors use 10~per~cent of \Rtwohun\ \citep[e.g.,][]{Ceverino2014} or a multiple of the half-mass radius \citep[e.g.,][]{Genel2014}.
}
\citet{Stevens2014} tested different methods defining the galactic edge in simulations and, while they advise against using a fraction of the virial radius, they found that the chosen technique does not significantly affect comparisons with observed scaling relations.
It should also be highlighted that observations are not devoid of similar debates and different methods can lead to different results as well \citep[][]{Bernardi2013}.

It is apparent from Fig.~\ref{fig:surf_den_maps} that the direct surroundings of some galaxies is rather chaotic with the presence of close-by satellites about to merge.
We therefore compare the values we obtain for $M_\star$, when taking $0.1\,\Rtwohun$ with respect to our fiducial $0.2\,\Rtwohun$, and make sure that $M_\star$ does not vary by more than $\sim$10~per~cent at the final redshift.
We also control that the variation in \Reff, $\Sigma_\mathrm{e}$, and SFR also remains within these bounds.
The results are visible in the following Table~\ref{tab:appendix}.

\begin{table}
    \centering
    \begin{tabular}{lrrrrr}
        \toprule
        Name & $z_{\mathrm{final}}$ & $\Delta M_\star$ & $\Delta \Reff$ & $\Delta \Sigmae$ & $\Delta\mathrm{SFR}_{100}$\\
        \midrule
        g3.08e12 & 2.0 &  $-$1.2 & $-$1.4 &        0.9 &    $-$0.5\\
        g3.09e12 & 2.0 &  $-$2.2 &  0.1 &       $-$0.1 &    0.0\\
        g3.00e12 &2.0&  $-$4.1 & $-$2.4 &        1.8 &    $-$0.5\\
        g3.20e12 &2.0&  $-$2.2 & $-$1.6 &        1.8 &    $-$0.4\\
        g2.75e12 &2.0&  $-$1.8 & $-$2.3 &        2.8 &    $-$0.9\\
        g3.03e12 &2.0&  $-$2.2 & $-$1.7 &        2.0 &    0.0\\
        g3.01e12 &2.0&  $-$1.2 & $-$0.1 &        0.1 &    $-$0.2\\
        g2.29e12 &2.0&  $-$1.9 & $-$1.3 &        1.5 &    $-$0.3\\
        g3.35e12 &2.0&  $-$9.0 & $-$6.5 &        5.4 &    $-$0.9\\
        g3.31e12 &2.0&  $-$4.0 & $-$5.1 &        4.2 &    $-$2.0\\
        g3.25e12 &2.0&  $-$7.9 & $-$6.6 &        7.9 &    $-$4.8\\
        g3.38e12 &2.0&  $-$13.8 & $-$11.8 &       12.8 &    $-$1.6\\
        g3.36e12 &2.0&  $-$1.2 & 0.0 &        0.0 &    $-$0.6\\
        g2.83e12 &2.0& $-$10.6 & $-$10.0 &        9.6 &   $-$12.3\\
        g2.63e12 &2.0&  $-$1.5 & $-$0.1 &        0.1 &    0.0\\
        g3.04e12 &2.0&  $-$2.1 & $-$0.9 &        0.8 &    $-$0.1\\
        g2.91e12 &2.0&  $-$0.3 & $-$0.3 &        0.1 &    0.0\\
        \midrule
        g2.47e12 &3.6&  $-$3.9 & $-$1.6 &        1.2 &    $-$0.9\\
        g2.40e12 &3.6&  $-$5.5 & $-$4.1 &        3.2 &    $-$4.6\\
        g2.69e12 &3.6&  $-$9.1 & $-$6.3 &        5.7 &    $-$1.1\\
        g3.76e12 &3.6&  $-$8.1 & $-$4.1 &        3.4 &    $-$1.3\\
        \textbf{g4.36e12} &\textbf{3.6}& \textbf{$-$30.8} &\textbf{$-$39.1} &       \textbf{73.2} &   \textbf{$-$26.0}\\
        g4.58e12 &3.6&  $-$4.7 & $-$3.5 &        4.3 &    $-$0.9\\
        g2.71e12 &3.6&  $-$4.3 & $-$2.0 &        1.8 &    $-$0.9\\
        g2.49e12 &3.6&  $-$5.3 & $-$3.3 &        2.8 &    $-$1.2\\
        g2.51e12 &3.6&  $-$2.0 & $-$1.7 &        1.7 &    $-$0.9\\
        g7.37e12 &3.6&  $-$6.0 & $-$6.2 &        6.8 &    $-$2.2\\
        g2.32e12 &3.6&  $-$2.5 & $-$0.8 &        0.8 &    $-$0.4\\
        g2.96e12 &3.6& $-$11.7 & $-$9.9 &        9.0 &    $-$6.6\\
        g7.97e12 &3.6&  $-$4.8 & $-$7.7 &        7.8 &    $-$3.6\\
        g9.20e12 &3.6&  $-$6.4 & $-$4.5 &        5.6 &    $-$1.8\\
        \textbf{g9.85e12} &\textbf{3.6}& \textbf{$-$39.5} &\textbf{$-$64.4} &      \textbf{387.1} &   \textbf{$-$37.7}\\
        \bottomrule
    \end{tabular}
    \caption{
    Columns from left to right: galaxy name, final redshift, and percentage differences in $M_\star$, \Reff, \Sigmae, and SFR$_{100}$, when using $0.1\,\Rtwohun$ instead of the fiducial $0.2\,\Rtwohun$ as each galaxy's defined radius.
    The two galaxies in \textbf{bold} were not used in this work.
    }
    \label{tab:appendix}
\end{table}

\begin{figure*}
\centering
\includegraphics[clip=true, trim= 2mm 2mm 2mm 2mm, width=\linewidth]{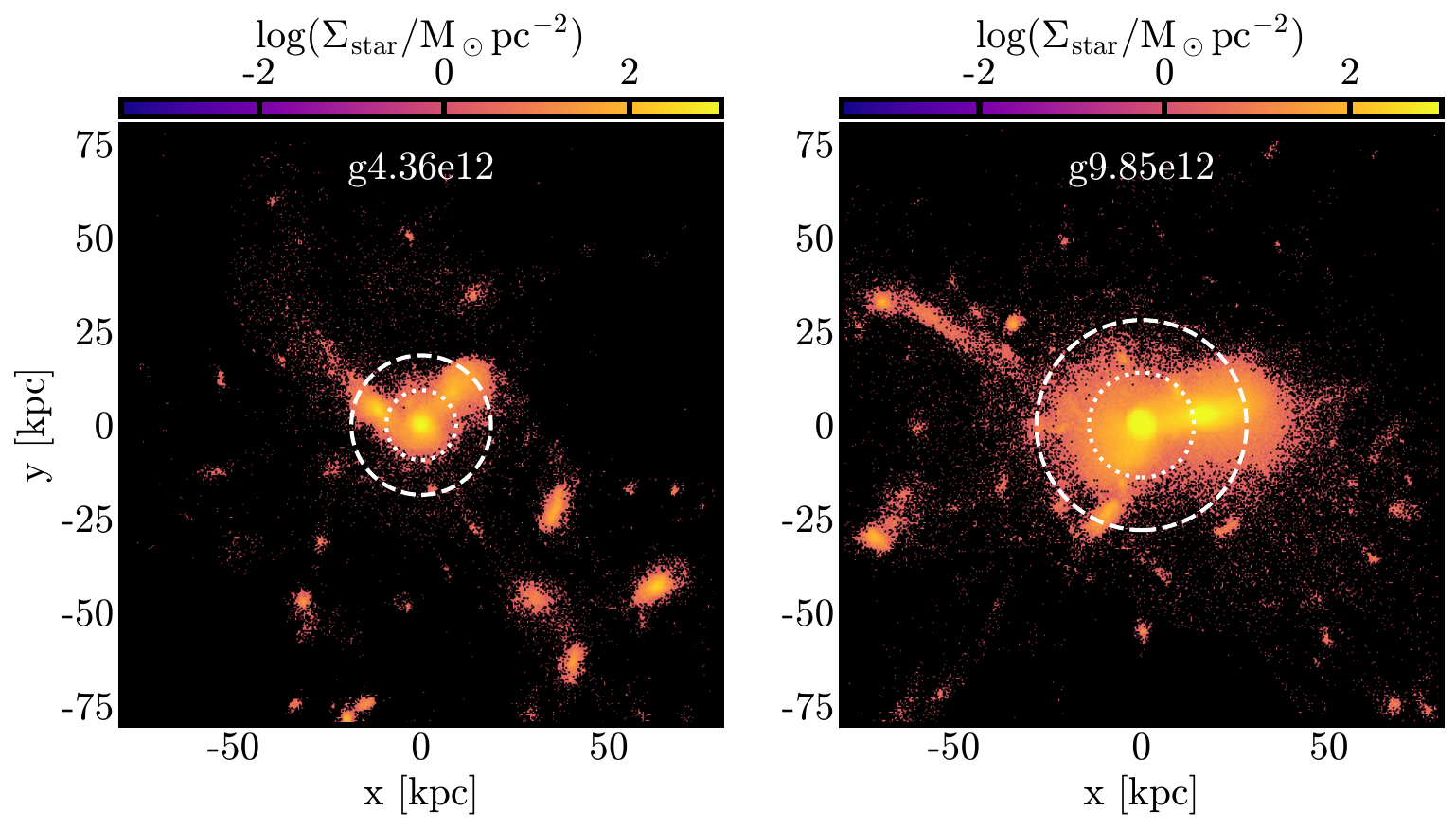}
\caption{
Stellar surface density maps viewed face-on for g4.36e12 and g9.85e12 that were excluded from this work.
The white dashed and dotted circles indicate a galactic radius of $0.2\,\Rtwohun$ and $0.1\,\Rtwohun$, respectively.
}
\label{fig:surf_den_maps_appendix}
\end{figure*}

We highlighted the two galaxies that we did not use in this paper due to the significant difference in $M_\star$ (and all other quantities) when reducing the galactic radius to $0.1\,\Rtwohun$.
For completeness, we show their stellar surface density maps in Fig.~\ref{fig:surf_den_maps_appendix}.
These maps are similar to Fig.~\ref{fig:surf_den_maps} where, in this case, the outer dashed and inner dotted circles represent $0.2\,\Rtwohun$ and $0.1\,\Rtwohun$, respectively.
As illustrated in Table~\ref{tab:appendix}, these two merger events significantly affected the calculated stellar quantities with respect to each galaxy's chosen edge.
We thus excluded them from this paper.

\section{Computing the energy from stellar feedback}\label{app:feedback_calculations}
In what follows we detail the methodology employed to calculate the energies associated with ESF and SNe feedback. We calculate both quantities separately during each snapshot before combining them to obtain the total stellar feedback energy.

The energy release during each simulation snapshot is calculated for stars formed within specific time intervals relative to the snapshots. For a given snapshot s at cosmic age $t$ in Myr, the relevant stars are those formed in the time interval $t_{\mathrm{s}-1} - 4 < t_\mathrm{form} < t_\mathrm{s} - 4$, where $t_\mathrm{form}$ is the age of the universe when the star particle was formed and the subtraction by 4 refers to the 4 Myr during which young stars release ESF and before which the eligible stars explode as SNe. For the calculation, we assume that both feedback energies are released at once 4 Myr after the stellar particle was formed.
In other words, if e.g., snapshot $i$ corresponds to $t=1$\,Gyr and snapshot $i+1$ to 1.2\,Gyr, star particles formed in the interval 0.96\,Gyr to 1.16\,Gyr will contribute to the feedback energy counted in snapshot $i+1$.

Once only the relevant stars are selected, the amount ESF ejected during one snapshot is calculated as:
\begin{equation}
    \Delta\mathrm{ESF} = \epsilon_\mathrm{ESF} \cdot 2\times10^{50}\,\mathrm{erg/\Msun}\, \cdot M_\mathrm{form},
\end{equation}
where $2\times10^{50}$ is the amount of energy in erg released per solar mass formed ($M_\mathrm{form}$) from SNe and $\epsilon_\mathrm{ESF} = 0.13$ is the fraction of this energy ejected as isotropic thermal energy.

The SN feedback energy is computed based on the number of stars within the mass range 8-40 \Msun\ that are expected to explode as SN\,II, expressed as $N_\mathrm{SN}$. This is determined by integrating the Chabrier IMF \citep[][]{Chabrier2003} over the specified mass range, scaled by the total $M_\mathrm{form}$. The SN feedback energy is then given by:
\begin{equation}
    \Delta\mathrm{SN} = \epsilon_\mathrm{SN} \cdot N_\mathrm{SN} \cdot 10^{51}\,\mathrm{erg/SN}.
\end{equation}

% Don't change these lines
\bsp	% typesetting comment
\label{lastpage}
\end{document}